\documentclass[lettersize,journal]{IEEEtran}
\usepackage{amsmath,amsfonts}
\usepackage{amssymb}
\usepackage{algorithmic}
\usepackage{array}
\usepackage[caption=false,font=normalsize,labelfont=sf,textfont=sf]{subfig}
\usepackage{textcomp}
\usepackage{stfloats}
\usepackage{url}
\usepackage{verbatim}
\usepackage{graphicx}
\def\BibTeX{{\rm B\kern-.05em{\sc i\kern-.025em b}\kern-.08em
    T\kern-.1667em\lower.7ex\hbox{E}\kern-.125emX}}
\usepackage{balance}
\begin{document}
\title{Performance Benchmarks for Line Spectral Estimation: Ordered Ziv--Zakai Characterization and Plug-In Amplitude Error Analysis}
\author{
Fangqing Xiao, \textit{Member, IEEE}, \quad
Dirk T. M. Slock, \textit{Life Fellow, IEEE}

\thanks{
Fangqing Xiao is with the School of Information Science and Engineering, 
Yunnan University, Kunming 650091, China (e-mail: fangqing.xiao@ynu.edu.cn).
}

\thanks{
Dirk T. M. Slock is with the Communication Systems Department, 
EURECOM, 06410 Biot, France (e-mail: dirk.slock@eurecom.fr).
}

\thanks{
Corresponding author: Fangqing Xiao (e-mail: fangqing.xiao@ynu.edu.cn).
}}


\maketitle

\begin{abstract}
Line spectral estimation (LSE) involves estimating both spectral frequencies
and their associated complex amplitudes. Existing Fisher-information-based
benchmarks are local and therefore do not capture either the threshold
behavior of frequency estimation or the propagation of frequency errors to
subsequent amplitude reconstruction. This paper develops explicit performance
benchmarks for LSE from two complementary perspectives: ordered frequency
estimation and plug-in amplitude reconstruction. On the frequency side, we
develop a computable Ziv--Zakai bound (ZZB)-type benchmark under an ordered
prior by combining a generalized-likelihood-ratio-test (GLRT)-based surrogate
for the unavailable pairwise kernel with an ordered-prior correction. The
resulting benchmark recovers the ordered a priori bound at low
signal-to-noise ratio (SNR) and the marginalized frequency-side
Cram{\'e}r--Rao bound (CRB) at high SNR. On the amplitude side, we derive a
local transfer characterization for the sequential plug-in estimator and
obtain a computable benchmark for the induced amplitude error. The resulting
framework explicitly characterizes threshold behavior on the frequency side
and error propagation on the amplitude side. Numerical results support the
proposed benchmarks across different SNR regimes, snapshot numbers, and model
orders.
\end{abstract}

\begin{IEEEkeywords}
Line spectral estimation, performance analysis, Ziv--Zakai bound,
Cram{\'e}r--Rao bound, amplitude error analysis.
\end{IEEEkeywords}

\section{Introduction}dd

\IEEEPARstart{L}{ine} spectral estimation (LSE) is a fundamental problem in
statistical signal processing and arises in many applications, including
radar, wireless communications, array processing, system identification, and
harmonic retrieval \cite{vantrees2002optimum}. Many estimation methods have
been developed for this problem, including maximum-likelihood and nonlinear
least-squares approaches
\cite{stoica1989musicmlcrb,Xunmeng2022Maximum,Babak2016Newtonized},
subspace-based methods
\cite{schmidt1986multiple,roy1989esprit,liu2015remarks}, sparse and gridless
methods \cite{malioutov2005sparse,zhou2018offgrid,Badiu2017Variational}, and
machine-learning-based algorithms
\cite{barthelme2021machine,papageorgiou2021deep,Xiao2025DoA, Xiao2026Multipath}. In many
practical settings, however, one is interested not only in locating the
spectral lines but also in recovering the associated complex amplitudes.
Consequently, a useful performance characterization should describe both the
accuracy of the frequency estimates themselves and the effect of frequency
uncertainty on the subsequent amplitude reconstruction stage.

Performance characterization is therefore as important as algorithm design.
Since line spectral estimation is a nonlinear parameter estimation problem, the
minimum mean-square error (MSE) is generally unavailable in closed form, which
motivates the use of performance bounds. The Cram{\'e}r--Rao bound (CRB) is
the most widely used local reference because it is determined by the Fisher
information matrix
\cite{vantrees2002optimum, liu2017crb,  cramer1946mathematical,kay1993fundamentals}. For
line spectral estimation, one may derive a joint Fisher-information-based
bound for the frequencies and amplitudes and then extract the corresponding
frequency-side and amplitude-side local references
\cite{stoica1989musicmlcrb,stoica1990performance}. However, CRB-type
characterizations are asymptotically tight only in the small-error regime and
do not capture the threshold behavior typical of line spectral estimation.
Global Bayesian bounds such as the Bayesian CRB
\cite{vantrees2007bayesian}, the Weiss--Weinstein bound
\cite{weiss1985lower,renaux2008fresh}, and the Bobrovsky--Zakai bound
\cite{bobrovsky1976lower} provide alternatives outside the asymptotic regime,
but they usually involve more demanding prior modeling, higher-order
derivatives, or nontrivial optimization. Moreover, when amplitudes are
reconstructed only after the frequencies have been estimated, the overall
performance is not determined solely by the frequency MSE, but also by how the
frequency error propagates through the reconstruction map.

Among global bounds, the Ziv--Zakai bound (ZZB) is particularly attractive
because it links the MSE to the minimum probability of error of an associated
binary hypothesis test and thus captures the transition from the a priori
region to the asymptotic region
\cite{ziv1969some,gu1991modified, seidman1970performance,chazan1975improved,bell1997extended}.
ZZB-type analyses have been developed for a broad range of nonlinear
estimation problems, including time-delay estimation
\cite{sadler2010ziv,decarli2014ziv,mishra2017performance,laas2021ziv,
zhang2022compressive, khan2013explicit}, bearing and direction-of-arrival estimation
\cite{khan2013explicit, bell1996explicit}, and multi-parameter radar estimation
\cite{chiriac2015ziv}. Recent ZZB results for multi-source one-dimensional and
two-dimensional direction-of-arrival estimation further show that, when the
performance metric depends on ordering or matching among multiple estimated
spectral parameters, the prior geometry must be handled carefully
\cite{Zhang2023ZZBdoa,Zhang2024ZZB2D, deCarvalho2013DML}. These developments are closely related
to the present problem, but they do not directly apply here. The main
difficulties are threefold. First, the amplitude matrix in the present model
is unknown and deterministic, so the pairwise likelihood-ratio test (LRT)
underlying the ZZB becomes composite and is generally unavailable in closed
form; a generalized likelihood-ratio test is therefore a natural alternative
\cite{decarli2014ziv,gabriel2005relationship}. Second, even with a computable
surrogate, the resulting vector ZZB remains difficult to evaluate because of
the ordered-support integration and the constrained optimization over
perturbation vectors. In addition, because the metric of interest is the MSE
of the ordered frequency vector, the correct low-SNR behavior must preserve
the geometry induced by the ordered prior. Third, beyond frequency estimation
itself, one must quantify how frequency uncertainty affects the subsequent
plug-in reconstruction of the amplitude matrix. This amplitude-side question
is not merely auxiliary: in practical estimation pipelines, it determines how
much of the frequency-side uncertainty is inherited by the final signal
reconstruction stage.

This paper addresses these issues from two complementary but connected
perspectives: ordered frequency estimation and plug-in amplitude
reconstruction. On the frequency side, we develop an explicit ZZB-type
benchmark for the ordered frequency MSE under unknown deterministic
amplitudes. Since the canonical pairwise likelihood-ratio kernel is not
available in closed form in this setting, we replace it with a computable
generalized-likelihood-ratio-test (GLRT)-based surrogate and derive a
tractable componentwise characterization through a local/nonlocal
decomposition. The resulting explicit construction has the correct local
Fisher limit, but its prior-scale term does not fully preserve the ordered
low-SNR geometry. We therefore introduce an ordered-prior correction that
restores the proper ordered a priori limit while retaining consistency with
the marginalized frequency-side CRB at high SNR. On the amplitude side,
rather than pursuing a second global Bayesian characterization, we analyze
the practical sequential plug-in estimator used in LSE pipelines. In
particular, we derive a local transfer law that quantifies how frequency
estimation errors propagate through the least-squares reconstruction map and
thereby induce amplitude reconstruction error. This leads to a computable
amplitude-side benchmark that complements the frequency-side ZZB-type
characterization: the former captures threshold behavior in ordered spectral
localization, whereas the latter characterizes how the resulting frequency
uncertainty is inherited by the subsequent reconstruction stage.

The main contributions of this paper are as follows. First, we develop an
explicit ZZB-type benchmark for the ordered frequency MSE in line spectral
estimation with unknown deterministic amplitudes by replacing the unavailable
canonical pairwise kernel with a GLRT-based surrogate and deriving a
tractable computable expression. Second, we show that this explicit
construction is locally Fisher-consistent but does not recover the correct
ordered low-SNR prior limit, and we resolve this mismatch by introducing an
ordered-prior correction that restores the proper a priori asymptote without
altering the high-SNR behavior. Third, we derive a local transfer
characterization for the sequential plug-in amplitude estimator and obtain a
computable benchmark for the induced amplitude MSE, thereby linking
frequency-estimation accuracy to final amplitude-reconstruction performance.

The rest of this paper is organized as follows.
Section~\ref{sec:system_model} introduces the system model, the frequency-side
and amplitude-side MSE criteria, and the corresponding local Fisher
benchmarks. Section~\ref{sec:zzb_preliminaries} formulates the ZZB for the
ordered frequency MSE and identifies the main computational obstacles.
Section~\ref{sec:componentwise_zzb} develops a computable basic frequency
benchmark from a GLRT-based ZZB-type construction.
Section~\ref{sec:ordered_prior_correction} introduces the ordered-prior
correction and discusses the resulting asymptotic interpretation.
Section~\ref{sec:induced_x_benchmark} establishes the local transfer analysis
for the plug-in amplitude MSE and derives a computable amplitude-side
benchmark. Section~\ref{sec:simulation} presents simulation results, and
Section~\ref{sec:conclusion} concludes the paper.

\textit{Notation:} Lowercase and uppercase boldface letters denote vectors and
matrices, respectively. The superscripts $(\cdot)^T$ and $(\cdot)^H$ denote
transpose and conjugate transpose. $\mathbb{C}$ and $\mathbb{R}$ denote the
complex and real fields, respectively. $\operatorname{Tr}(\cdot)$,
$\operatorname{vec}(\cdot)$, $\mathbb{E}[\cdot]$, $\Re(\cdot)$, and $\|\cdot\|_F$ denote the
trace, vectorization, expectation, real part operation, and Frobenius norm, respectively.
$\operatorname{Diag}(\cdot)$ denotes a diagonal matrix formed from its
arguments, $\odot$ denotes the Hadamard product, and
$\mathcal{CN}(\cdot,\cdot)$ denotes the circularly symmetric complex Gaussian
distribution. In addition, $Q(\cdot)$, $\phi(\cdot)$, and $\Phi(\cdot)$ denote
the Gaussian tail function, the standard normal density, and the standard
normal cumulative distribution function, respectively.
\section{System Model and Local Fisher Benchmarks}
\label{sec:system_model}

This section introduces the signal model and the local Fisher references used throughout the paper, treating both the frequency vector and the amplitude matrix as unknown deterministic parameters.

Consider the $K$-tone line spectral model observed over $T$ snapshots:
\begin{equation}
\mathbf{Y}
=
\mathbf{A}(\boldsymbol{\omega})\mathbf{X}
+
\mathbf{N},
\label{eq:model}
\end{equation}
where $\mathbf{Y}\in\mathbb{C}^{M\times T}$ is the observation matrix,
$\mathbf{X}\in\mathbb{C}^{K\times T}$ is an unknown deterministic amplitude
matrix, and $\mathbf{N}\in\mathbb{C}^{M\times T}$ is additive noise. The
unknown frequency vector is
\begin{equation}
\boldsymbol{\omega}
=
[\omega_1,\omega_2,\ldots,\omega_K]^T,
\label{eq:omega_def}
\end{equation}
and the associated Vandermonde matrix is
\begin{equation}
\mathbf{A}(\boldsymbol{\omega})
=
[\mathbf{a}(\omega_1),\mathbf{a}(\omega_2),\ldots,\mathbf{a}(\omega_K)]
\in\mathbb{C}^{M\times K},
\label{eq:A_def}
\end{equation}
with steering vector
\begin{equation}
\mathbf{a}(\omega_k)
=
[1,e^{j\omega_k},\ldots,e^{j(M-1)\omega_k}]^T.
\label{eq:steering_def}
\end{equation}
Throughout the paper, the frequencies are assumed to satisfy
\begin{equation}
\omega_{\min}\le \omega_1<\omega_2<\cdots<\omega_K\le \omega_{\max}.
\label{eq:ordered_support}
\end{equation}
Define
\begin{equation}
\zeta \triangleq \omega_{\max}-\omega_{\min},
\label{eq:zeta_def}
\end{equation}
and the ordered support
\begin{equation}
\Omega_{\mathrm{ord}}
=
\left\{
\boldsymbol{\omega}:
\omega_{\min}\le \omega_1<\cdots<\omega_K\le \omega_{\max}
\right\}.
\label{eq:Omega_ord}
\end{equation}

We adopt the following assumptions.

\textit{Assumption 1:}
The entries of $\mathbf{N}$ are i.i.d.\ circularly symmetric complex Gaussian
random variables distributed as $\mathcal{CN}(0,\sigma^2)$.

\textit{Assumption 2:}
The frequency components are distinct and ordered as in
\eqref{eq:ordered_support}.

\textit{Assumption 3:}
The matrix $\mathbf{A}(\boldsymbol{\omega})$ has full column rank in the
parameter region of interest.

\textit{Assumption 4:}
The ordered prior over $\Omega_{\mathrm{ord}}$ is used only in the subsequent
global ZZB analysis. In the present section, both $\boldsymbol{\omega}$ and
$\mathbf{X}$ are treated as unknown deterministic parameters.

Assumptions 1--3 ensure that the local Fisher quantities introduced below are
well defined, whereas Assumption 4 separates the present deterministic
formulation from the later ordered-prior analysis.

Let $\hat{\boldsymbol{\omega}}$ and $\hat{\mathbf{X}}$ denote the estimators of
$\boldsymbol{\omega}$ and $\mathbf{X}$, respectively. The frequency-side metric
of interest is the average mean-square error (MSE)
\begin{equation}
\mathbf{R}_{\boldsymbol{\omega}}
=
\mathbb{E}
\!\left[
(\hat{\boldsymbol{\omega}}-\boldsymbol{\omega})
(\hat{\boldsymbol{\omega}}-\boldsymbol{\omega})^T
\right],
\label{eq:Romega_def}
\end{equation}
\begin{equation}
\mathrm{MSE}_{\boldsymbol{\omega}}
=
\frac{1}{K}\operatorname{Tr}(\mathbf{R}_{\boldsymbol{\omega}}).
\label{eq:MSEomega_def}
\end{equation}
For the amplitude matrix, we use the normalized Frobenius MSE
\begin{equation}
\mathrm{MSE}_{\mathbf{X}}
=
\frac{1}{KT}
\mathbb{E}
\!\left[
\|\hat{\mathbf{X}}-\mathbf{X}\|_F^2
\right].
\label{eq:MSEX_def}
\end{equation}
These two metrics play different roles in the sequel: the frequency-side MSE
will be characterized globally through a ZZB-type construction, whereas the
amplitude-side MSE will later be analyzed through a local transfer law for
the plug-in estimator.

We next introduce the local Fisher information for the deterministic-parameter
formulation. Write
\begin{equation}
\mathbf{X}
=
\mathbf{X}_{\mathrm{R}}
+
j\mathbf{X}_{\mathrm{I}},
\label{eq:X_real_imag}
\end{equation}
where $\mathbf{X}_{\mathrm{R}},\mathbf{X}_{\mathrm{I}}\in\mathbb{R}^{K\times T}$,
and define the real-valued parameter vector
\begin{equation}
\boldsymbol{\eta}
=
\begin{bmatrix}
\boldsymbol{\omega}^T &
\operatorname{vec}(\mathbf{X}_{\mathrm{R}})^T &
\operatorname{vec}(\mathbf{X}_{\mathrm{I}})^T
\end{bmatrix}^T.
\label{eq:eta_def}
\end{equation}
Let
\begin{equation}
\mathbf{x}
\triangleq
\begin{bmatrix}
\operatorname{vec}(\mathbf{X}_{\mathrm{R}})^T &
\operatorname{vec}(\mathbf{X}_{\mathrm{I}})^T
\end{bmatrix}^T,
\label{eq:x_def}
\end{equation}
and partition the Fisher information matrix (FIM) with respect to
$\boldsymbol{\eta}$ as
\begin{equation}
\mathbf{J}_{\boldsymbol{\eta}}
=
\begin{bmatrix}
\mathbf{J}_{\boldsymbol{\omega}\boldsymbol{\omega}}
&
\mathbf{J}_{\boldsymbol{\omega}\mathbf{x}}
\\[1mm]
\mathbf{J}_{\mathbf{x}\boldsymbol{\omega}}
&
\mathbf{J}_{\mathbf{x}\mathbf{x}}
\end{bmatrix}.
\label{eq:FIM_partition}
\end{equation}
The corresponding CRB matrix is
\begin{equation}
\mathbf{C}_{\boldsymbol{\eta}}
=
\mathbf{J}_{\boldsymbol{\eta}}^{-1}.
\label{eq:Ceta_def}
\end{equation}

For frequency estimation, $\mathbf{X}$ acts as a nuisance parameter. The
relevant local reference is therefore not the conditional CRB under known
amplitudes, but the marginalized frequency-side CRB obtained after eliminating
$\mathbf{X}$. It is given by the
$\boldsymbol{\omega}$-block of $\mathbf{C}_{\boldsymbol{\eta}}$:
\begin{equation}
\mathbf{C}_{\boldsymbol{\omega}}
=
\bigl[\mathbf{C}_{\boldsymbol{\eta}}\bigr]_{\boldsymbol{\omega}\boldsymbol{\omega}}
=
\Bigl(
\mathbf{J}_{\boldsymbol{\omega}\boldsymbol{\omega}}
-
\mathbf{J}_{\boldsymbol{\omega}\mathbf{x}}
\mathbf{J}_{\mathbf{x}\mathbf{x}}^{-1}
\mathbf{J}_{\mathbf{x}\boldsymbol{\omega}}
\Bigr)^{-1}.
\label{eq:Comega_def}
\end{equation}
This matrix serves as the local asymptotic benchmark for the subsequent
frequency-side analysis.

To make the corresponding Fisher geometry explicit, define
\begin{equation}
\mathbf{P}_{\mathbf A}
=
\mathbf{A}(\boldsymbol{\omega})
\bigl(
\mathbf{A}^H(\boldsymbol{\omega})\mathbf{A}(\boldsymbol{\omega})
\bigr)^{-1}
\mathbf{A}^H(\boldsymbol{\omega}),
\label{eq:PA_def}
\end{equation}
\begin{equation}
\mathbf{P}_{\mathbf A}^{\perp}
=
\mathbf{I}_M-\mathbf{P}_{\mathbf A},
\label{eq:PAperp_def}
\end{equation}
and
\begin{equation}
\mathbf{D}_i
\triangleq
\frac{\partial \mathbf{A}(\boldsymbol{\omega})}{\partial \omega_i},
\qquad i=1,\ldots,K.
\label{eq:Di_def}
\end{equation}
The equivalent Fisher information matrix for $\boldsymbol{\omega}$ after
eliminating $\mathbf{X}$, that is, the Schur complement of the joint FIM with
respect to the nuisance parameter $\mathbf{x}$, can be written for the present
model in the projector form \cite{kay1993fundamentals}
\begin{equation}
\bigl[\mathbf{J}_{\boldsymbol{\omega}}^{\mathrm{eff}}\bigr]_{ij}
=
\frac{2}{\sigma^2}
\Re\!\left\{
\operatorname{Tr}
\!\left(
\mathbf{X}^H
\mathbf{D}_i^H
\mathbf{P}_{\mathbf A}^{\perp}
\mathbf{D}_j
\mathbf{X}
\right)
\right\},
\qquad i,j=1,\ldots,K,
\label{eq:Jw_eff}
\end{equation}
so that
\begin{equation}
\mathbf{C}_{\boldsymbol{\omega}}
=
\bigl(
\mathbf{J}_{\boldsymbol{\omega}}^{\mathrm{eff}}
\bigr)^{-1}.
\label{eq:Comega_eff}
\end{equation}
This expression makes clear that the local frequency geometry is determined
after projecting out the amplitude nuisance space.

For later use, the derivative of the steering vector is
\begin{equation}
\mathbf{a}'(\omega_i)
=
j
\begin{bmatrix}
0 & 1 & \cdots & M-1
\end{bmatrix}^T
\odot
\mathbf{a}(\omega_i),
\label{eq:aprime_def}
\end{equation}
so that $\mathbf{D}_i$ contains $\mathbf{a}'(\omega_i)$ in its $i$th column and
zeros elsewhere.

Equations \eqref{eq:Ceta_def}--\eqref{eq:Comega_eff} provide the local Fisher
references used in the sequel, in particular the marginalized frequency-side
CRB that will serve as the local benchmark for the later ZZB-type analysis.
\section{ZZB Formulation and Computational Obstacles}
\label{sec:zzb_preliminaries}
To characterize the global performance of the ordered frequency vector
$\boldsymbol{\omega}$, we start from the extended Ziv--Zakai bound (ZZB) for
vector parameter estimation. For any unit-norm vector
$\boldsymbol{z}\in\mathbb{R}^K$ \cite[Eq.~(32)]{bell1997extended},
\begin{equation}
\boldsymbol{z}^T\mathbf{R}_{\boldsymbol{\omega}}\boldsymbol{z}
\ge
\frac{1}{2}
\int_0^\infty
h\,
\max_{\boldsymbol{\delta}:\,\boldsymbol{z}^T\boldsymbol{\delta}=h}
\Psi(\boldsymbol{\delta})
\,dh,
\label{eq:vzzb_standard_compact}
\end{equation}
where
\begin{equation}
\Psi(\boldsymbol{\delta})
\triangleq
\int_{\mathbb{R}^K}
\bigl(
f_{\boldsymbol{\omega}}(\boldsymbol{\phi})
+
f_{\boldsymbol{\omega}}(\boldsymbol{\phi}+\boldsymbol{\delta})
\bigr)
P_{\min}(\boldsymbol{\phi},\boldsymbol{\phi}+\boldsymbol{\delta})
\,d\boldsymbol{\phi},
\label{eq:Psi_def}
\end{equation}
$f_{\boldsymbol{\omega}}(\cdot)$ is the prior density of
$\boldsymbol{\omega}$, and
$P_{\min}(\boldsymbol{\phi},\boldsymbol{\phi}+\boldsymbol{\delta})$
denotes the minimum probability of error of the associated binary hypothesis
test.

For the model in \eqref{eq:model}, the pairwise hypotheses associated with
$\boldsymbol{\phi}$ and $\boldsymbol{\phi}+\boldsymbol{\delta}$ are
\begin{equation}
\mathcal{H}_0:\ \boldsymbol{\omega}=\boldsymbol{\phi},
\qquad
\mathcal{H}_1:\ \boldsymbol{\omega}=\boldsymbol{\phi}+\boldsymbol{\delta},
\label{eq:pairwise_hypotheses}
\end{equation}
with prior probabilities
\begin{equation}
\Pr(\mathcal{H}_0)
=
\frac{
f_{\boldsymbol{\omega}}(\boldsymbol{\phi})
}{
f_{\boldsymbol{\omega}}(\boldsymbol{\phi})
+
f_{\boldsymbol{\omega}}(\boldsymbol{\phi}+\boldsymbol{\delta})
},
\qquad
\Pr(\mathcal{H}_1)=1-\Pr(\mathcal{H}_0).
\label{eq:pairwise_priors}
\end{equation}
Under the two hypotheses,
\begin{equation}
\mathcal{H}_0:\ \mathbf{Y}=\mathbf{A}(\boldsymbol{\phi})\mathbf{X}+\mathbf{N},
\qquad
\mathcal{H}_1:\ \mathbf{Y}=\mathbf{A}(\boldsymbol{\phi}+\boldsymbol{\delta})\mathbf{X}+\mathbf{N},
\label{eq:pairwise_observation_model}
\end{equation}
where $\mathbf{X}$ is unknown deterministic under both hypotheses and
$\mathbf{N}$ has i.i.d.\ entries distributed as $\mathcal{CN}(0,\sigma^2)$.
Equivalently,
\begin{equation}
\operatorname{vec}(\mathbf{Y})\mid \mathcal{H}_0
\sim
\mathcal{CN}\!\bigl(
\operatorname{vec}(\mathbf{A}(\boldsymbol{\phi})\mathbf{X}),
\ \sigma^2\mathbf{I}_{MT}
\bigr),
\label{eq:Y_dist_H0}
\end{equation}
and
\begin{equation}
\operatorname{vec}(\mathbf{Y})\mid \mathcal{H}_1
\sim
\mathcal{CN}\!\bigl(
\operatorname{vec}(\mathbf{A}(\boldsymbol{\phi}+\boldsymbol{\delta})\mathbf{X}),
\ \sigma^2\mathbf{I}_{MT}
\bigr).
\label{eq:Y_dist_H1}
\end{equation}
In the standard ZZB construction,
$P_{\min}(\boldsymbol{\phi},\boldsymbol{\phi}+\boldsymbol{\delta})$
is the minimum probability of error of the corresponding Bayesian binary test.

In the present paper, however, the quantity of interest is not a generic
directional quadratic form but the ordered frequency average MSE,
\begin{equation}
\mathrm{MSE}_{\boldsymbol{\omega}}
=
\frac{1}{K}\operatorname{Tr}(\mathbf{R}_{\boldsymbol{\omega}})
=
\frac{1}{K}\sum_{i=1}^K
\boldsymbol{e}_i^T
\mathbf{R}_{\boldsymbol{\omega}}
\boldsymbol{e}_i,
\label{eq:mse_trace_decomp_sec3}
\end{equation}
where $\boldsymbol{e}_i$ is the $i$th canonical basis vector in
$\mathbb{R}^K$. This representation is important for what follows. Although
\eqref{eq:vzzb_standard_compact} is stated for a generic direction
$\boldsymbol{z}$, the performance criterion of interest here is the trace
average of the ordered coordinate errors. The relevant construction is
therefore naturally componentwise: we specialize
\eqref{eq:vzzb_standard_compact} to $\boldsymbol{z}=\boldsymbol{e}_i$, derive
a characterization for each term
$\boldsymbol{e}_i^T\mathbf{R}_{\boldsymbol{\omega}}\boldsymbol{e}_i$, and then
average over $i$.

At this point, the main issue is no longer the formal ZZB expression itself,
but whether the associated pairwise kernel and ordered-support optimization
can be handled explicitly. Direct evaluation remains difficult for two
distinct reasons.

\textit{First}, the canonical pairwise kernel is unavailable in closed form.
In many Bayesian settings, nuisance quantities can be removed by prior
marginalization, so the pairwise test remains simple. Here, by contrast,
$\mathbf{X}$ is an unknown deterministic nuisance parameter. As a result, the
binary test induced by \eqref{eq:pairwise_observation_model} is composite
under both hypotheses, and the LRT-based kernel
$P_{\min}(\boldsymbol{\phi},\boldsymbol{\phi}+\boldsymbol{\delta})$
is not explicitly available. Thus, the first obstacle is statistical: the
canonical ZZB kernel is not directly computable under the present deterministic
amplitude model.

\textit{Second}, even after replacing the unavailable kernel by a computable
surrogate, the ZZB still involves an ordered-support integral together with a
constrained optimization over $\boldsymbol{\delta}$. In particular, after
restricting the prior to $\Omega_{\mathrm{ord}}$, the integral in
\eqref{eq:Psi_def} takes the form
\begin{align}
&\int_{\mathbb{R}^K}
\bigl(
f_{\boldsymbol{\omega}}(\boldsymbol{\phi})
+
f_{\boldsymbol{\omega}}(\boldsymbol{\phi}+\boldsymbol{\delta})
\bigr)
P(\boldsymbol{\phi},\boldsymbol{\delta})
\,d\boldsymbol{\phi}
\nonumber\\
&\qquad=
\frac{2}{|\Omega_{\mathrm{ord}}|}
\int_{\Omega_{\mathrm{ord}}\cap(\Omega_{\mathrm{ord}}-\boldsymbol{\delta})}
P(\boldsymbol{\phi},\boldsymbol{\delta})
\,d\boldsymbol{\phi},
\label{eq:ordered_overlap_integral}
\end{align}
subject to
\begin{equation}
\boldsymbol{e}_i^T\boldsymbol{\delta}=h
\qquad
(\text{or, more generally, } \boldsymbol{z}^T\boldsymbol{\delta}=h).
\label{eq:delta_constraint}
\end{equation}
For $K>1$, neither the overlap integral in
\eqref{eq:ordered_overlap_integral} nor the constrained maximization in
\eqref{eq:delta_constraint} is tractable in useful closed form. Thus, the
second obstacle is geometric: even with a computable surrogate kernel, the
ordered prior induces a coupled integration-and-optimization structure that
does not simplify directly.

\section{Computable Componentwise ZZB for Ordered Frequency MSE}
\label{sec:componentwise_zzb}

This section develops a computable componentwise ZZB-type benchmark for the
ordered frequency MSE. We first replace the unavailable pairwise kernel by a
GLRT-based surrogate, then introduce a local/nonlocal decomposition, and
finally obtain the explicit coordinate-wise benchmark and its averaged form.

\subsection{GLRT-Based Surrogate for the Pairwise Kernel}
\label{subsec:glrt_surrogate}

Fix $i\in\{1,\ldots,K\}$ and set $\boldsymbol{z}=\boldsymbol{e}_i$ in
\eqref{eq:vzzb_standard_compact}. The corresponding coordinate-wise ZZB is
\begin{equation}
\boldsymbol{e}_i^T\mathbf{R}_{\boldsymbol{\omega}}\boldsymbol{e}_i
\ge
\frac{1}{2}
\int_0^\infty
h\,
\max_{\boldsymbol{\delta}:\,\delta_i=h}
\Psi_i(\boldsymbol{\delta})\,dh,
\label{eq:zzb_ei_start}
\end{equation}
where
\begin{equation}
\Psi_i(\boldsymbol{\delta})
\triangleq
\int_{\mathbb{R}^K}
\bigl(
f_{\boldsymbol{\omega}}(\boldsymbol{\phi})
+
f_{\boldsymbol{\omega}}(\boldsymbol{\phi}+\boldsymbol{\delta})
\bigr)
P_{\min}(\boldsymbol{\phi},\boldsymbol{\phi}+\boldsymbol{\delta})
\,d\boldsymbol{\phi}.
\label{eq:Psi_i_def}
\end{equation}

For the pair $(\boldsymbol{\phi},\boldsymbol{\phi}+\boldsymbol{\delta})$, the
observation model is
\begin{equation}
\mathcal{H}_0:\ \mathbf{Y}=\mathbf{A}(\boldsymbol{\phi})\mathbf{X}+\mathbf{N},
\qquad
\mathcal{H}_1:\ \mathbf{Y}=\mathbf{A}(\boldsymbol{\phi}+\boldsymbol{\delta})\mathbf{X}+\mathbf{N},
\label{eq:pairwise_model_hyp}
\end{equation}
where $\mathbf{X}$ is unknown deterministic under both hypotheses. Hence the
pairwise test is composite, and the canonical LRT kernel
$P_{\min}(\boldsymbol{\phi},\boldsymbol{\phi}+\boldsymbol{\delta})$ is not
available in closed form. If one were to assign a prior to $\mathbf{X}$ and
marginalize it out, the resulting test would no longer correspond to the
deterministic amplitude model adopted in this paper. We therefore follow the
standard deterministic-nuisance route and replace $\mathbf{X}$ under each
hypothesis by its maximum-likelihood estimate, equivalently its least-squares
estimate under Gaussian noise. This leads naturally to a GLRT-based surrogate
for the unavailable canonical pairwise kernel.

For any candidate frequency vector $\boldsymbol{\nu}$, define the orthogonal
projector onto the column space of $\mathbf{A}(\boldsymbol{\nu})$ by
\begin{equation}
\mathbf{P}_{\boldsymbol{\nu}}
=
\mathbf{A}(\boldsymbol{\nu})
\bigl(
\mathbf{A}^H(\boldsymbol{\nu})\mathbf{A}(\boldsymbol{\nu})
\bigr)^{-1}
\mathbf{A}^H(\boldsymbol{\nu}).
\label{eq:projector_def}
\end{equation}
The corresponding least-squares estimate of $\mathbf{X}$ is
\begin{equation}
\hat{\mathbf{X}}(\boldsymbol{\nu})
=
\bigl(
\mathbf{A}^H(\boldsymbol{\nu})\mathbf{A}(\boldsymbol{\nu})
\bigr)^{-1}
\mathbf{A}^H(\boldsymbol{\nu})\mathbf{Y}.
\label{eq:Xhat_nu_sec4}
\end{equation}
Substituting \eqref{eq:Xhat_nu_sec4} into the residual energy yields
\begin{equation}
\|\mathbf{Y}-\mathbf{A}(\boldsymbol{\nu})\hat{\mathbf{X}}(\boldsymbol{\nu})\|_F^2
=
\|\mathbf{Y}\|_F^2
-
\operatorname{Tr}\!\left(
\mathbf{Y}^H\mathbf{P}_{\boldsymbol{\nu}}\mathbf{Y}
\right).
\label{eq:residual_projector_identity}
\end{equation}
Therefore, comparing the two least-squares residuals is equivalent to
comparing the projected energies onto the two candidate signal subspaces. The
resulting GLRT detector is
\begin{equation}
\Gamma(\mathbf{Y};\boldsymbol{\phi},\boldsymbol{\delta})
=
\operatorname{Tr}
\!\left(
\mathbf{Y}^H
\bigl(
\mathbf{P}_{\boldsymbol{\phi}}-\mathbf{P}_{\boldsymbol{\phi}+\boldsymbol{\delta}}
\bigr)
\mathbf{Y}
\right),
\label{eq:glrt_detector}
\end{equation}
and the associated pairwise kernel is
\begin{equation}
P_e(\boldsymbol{\phi},\boldsymbol{\delta})
=
\Pr\!\left(
\Gamma(\mathbf{Y};\boldsymbol{\phi},\boldsymbol{\delta})<0
\mid
\mathcal{H}_0
\right).
\label{eq:glrt_kernel_def}
\end{equation}

Let
\begin{equation}
\boldsymbol{\Delta}(\boldsymbol{\phi},\boldsymbol{\delta})
\triangleq
\mathbf{P}_{\boldsymbol{\phi}}
-
\mathbf{P}_{\boldsymbol{\phi}+\boldsymbol{\delta}}.
\label{eq:Delta_def}
\end{equation}
Under $\mathcal{H}_0$, write
\begin{equation}
\mathbf{M}_0
=
\mathbf{A}(\boldsymbol{\phi})\mathbf{X},
\qquad
\mathbf{Y}
=
\mathbf{M}_0+\mathbf{N}.
\label{eq:M0_def}
\end{equation}
Then
\begin{align}
\Gamma
&=
\operatorname{Tr}
\!\left(
(\mathbf{M}_0+\mathbf{N})^H
\boldsymbol{\Delta}
(\mathbf{M}_0+\mathbf{N})
\right)
\nonumber\\
&=
\operatorname{Tr}(\mathbf{M}_0^H\boldsymbol{\Delta}\mathbf{M}_0)
+
2\Re\!\left\{
\operatorname{Tr}(\mathbf{N}^H\boldsymbol{\Delta}\mathbf{M}_0)
\right\}
+
\operatorname{Tr}(\mathbf{N}^H\boldsymbol{\Delta}\mathbf{N}).
\label{eq:Gamma_expand}
\end{align}

Because $\mathbf{Y}=[\mathbf{y}_1,\ldots,\mathbf{y}_T]$ contains $T$
independent snapshots,
\begin{equation}
\Gamma
=
\sum_{t=1}^T \gamma_t,
\qquad
\gamma_t
=
\mathbf{y}_t^H
\boldsymbol{\Delta}
\mathbf{y}_t.
\label{eq:Gamma_snapshot_sum}
\end{equation}
Each $\gamma_t$ is an indefinite Hermitian quadratic form. Rather than work
with the exact finite-$T$ distribution of $\Gamma$, we approximate the
aggregated statistic by a Gaussian random variable using a central-limit-type
argument, matching its first two moments under $\mathcal{H}_0$. This
approximation is used only to obtain a computable ZZB-type benchmark.

Under $\mathcal{H}_0$, the conditional mean is
\begin{equation}
\mu_\Gamma(\boldsymbol{\phi},\boldsymbol{\delta})
=
\mathbb{E}[\Gamma\mid\mathcal{H}_0]
=
\operatorname{Tr}
\!\left(
\mathbf{M}_0^H\boldsymbol{\Delta}\mathbf{M}_0
\right),
\label{eq:muGamma}
\end{equation}
and the conditional variance is
\begin{equation}
\sigma_\Gamma^2(\boldsymbol{\phi},\boldsymbol{\delta})
=
2\sigma^2
\operatorname{Tr}
\!\left(
\mathbf{M}_0^H\boldsymbol{\Delta}^2\mathbf{M}_0
\right)
+
T\sigma^4\operatorname{Tr}(\boldsymbol{\Delta}^2),
\label{eq:sigmaGamma}
\end{equation}
where standard second-moment identities for complex Gaussian quadratic forms
have been used. We then write
\begin{equation}
\Gamma
\sim
\mathcal{N}
\bigl(
\mu_\Gamma(\boldsymbol{\phi},\boldsymbol{\delta}),
\sigma_\Gamma^2(\boldsymbol{\phi},\boldsymbol{\delta})
\bigr),
\label{eq:Gamma_gaussian}
\end{equation}
which gives the Gaussianized GLRT kernel
\begin{equation}
P_e(\boldsymbol{\phi},\boldsymbol{\delta})
\approx
Q\!\left(
\frac{
\mu_\Gamma(\boldsymbol{\phi},\boldsymbol{\delta})
}{
\sigma_\Gamma(\boldsymbol{\phi},\boldsymbol{\delta})
}
\right).
\label{eq:glrt_kernel_gaussian}
\end{equation}
This Gaussianized kernel replaces the unavailable canonical pairwise kernel
and provides the starting point for the explicit construction below.

\subsection{Local/Nonlocal Decomposition of the Coordinate-Wise ZZB}
\label{subsec:local_nonlocal_decomp}

Substituting \eqref{eq:glrt_kernel_gaussian} into
\eqref{eq:zzb_ei_start} yields
\begin{equation}
\boldsymbol{e}_i^T\mathbf{R}_{\boldsymbol{\omega}}\boldsymbol{e}_i
\gtrsim
\frac{1}{2}
\int_0^\infty
h\,
\max_{\boldsymbol{\delta}:\,\delta_i=h}
\widetilde{\Psi}_i(\boldsymbol{\delta})
\,dh,
\label{eq:zzb_ei_surrogate}
\end{equation}
where
\begin{equation}
\widetilde{\Psi}_i(\boldsymbol{\delta})
\triangleq
\int_{\mathbb{R}^K}
\bigl(
f_{\boldsymbol{\omega}}(\boldsymbol{\phi})
+
f_{\boldsymbol{\omega}}(\boldsymbol{\phi}+\boldsymbol{\delta})
\bigr)
P_e(\boldsymbol{\phi},\boldsymbol{\delta})
\,d\boldsymbol{\phi}.
\label{eq:Psi_tilde_def}
\end{equation}
At this point the detection kernel is explicit, but the
ordered-support integral and the constrained maximization remain analytically
coupled.

The key simplification is that the dominant mechanism changes with the scale
of $\boldsymbol{\delta}$. For small perturbations, the Gaussianized kernel is
governed by the local Fisher geometry after eliminating the nuisance
parameter $\mathbf{X}$. For larger perturbations, the local curvature no
longer dominates, and the kernel approaches a prior-scale plateau associated
with weak coupling between the two signal subspaces. This motivates a
local/nonlocal decomposition of the coordinate-wise benchmark.

To formalize this decomposition, let
\begin{equation}
P_{S,i}^\star(h)
\triangleq
\max_{\boldsymbol{\delta}:\,\delta_i=h}
P_S(\boldsymbol{\delta}),
\label{eq:PSi_star_def}
\end{equation}
where $P_S(\boldsymbol{\delta})$ denotes the local Fisher-based surrogate
introduced below in \eqref{eq:PS_local}, and let $\bar P_L$ denote the
representative nonlocal plateau level introduced later in
\eqref{eq:PLbar_again}. We then define a transition point $\tilde h_i$ through
the crossing rule
\begin{equation}
P_{S,i}^\star(\tilde h_i)=\bar P_L.
\label{eq:crossing_rule_def}
\end{equation}
The local contribution is taken over $0\le h\le \tilde h_i$, while the
remaining prior-scale contribution is assigned to the nonlocal term:
\begin{equation}
B_{i,\mathrm{L}}
\triangleq
\int_0^{\tilde h_i}
h\,
\bigl[
P_{S,i}^\star(h)-\bar P_L
\bigr]dh,
\label{eq:BiL_def}
\end{equation}
\begin{equation}
B_{i,\mathrm{NL}}
\triangleq
\frac{\bar P_L}{2}
\int_0^\zeta
h
\int_0^{\zeta-h}
\bigl[
f_i(u)+f_i(u+h)
\bigr]\,du\,dh.
\label{eq:BiNL_def}
\end{equation}
Accordingly,
\begin{equation}
\frac{1}{2}
\int_0^\infty
h\,
\max_{\boldsymbol{\delta}:\,\delta_i=h}
\widetilde{\Psi}_i(\boldsymbol{\delta})
\,dh
\approx
B_{i,\mathrm{L}}+B_{i,\mathrm{NL}}.
\label{eq:decomp_main}
\end{equation}
The subtraction of $\bar P_L$ in \eqref{eq:BiL_def} removes the plateau part
from the local integral, whereas \eqref{eq:BiNL_def} retains the corresponding
prior-scale contribution. The local term yields the Fisher-consistent
component of the benchmark, while the nonlocal term captures its prior-scale
behavior.

\subsection{Local Contribution}
\label{subsec:local_contribution}

In the local regime,
$\mathcal{R}(\mathbf{A}(\boldsymbol{\phi}))$ and
$\mathcal{R}(\mathbf{A}(\boldsymbol{\phi}+\boldsymbol{\delta}))$ remain close,
so the Gaussianized kernel is governed by the local Fisher geometry after
eliminating $\mathbf{X}$. We therefore use the approximation
\begin{equation}
P_S(\boldsymbol{\delta})
=
Q\!\left(
\frac{1}{2}
\sqrt{
\boldsymbol{\delta}^T
\mathbf{J}_{\boldsymbol{\omega}}^{\mathrm{eff}}
\boldsymbol{\delta}}
\right),
\label{eq:PS_local}
\end{equation}
where $\mathbf{J}_{\boldsymbol{\omega}}^{\mathrm{eff}}$ is defined in
\eqref{eq:Jw_eff}.

For the $i$th coordinate-wise term, the constraint is
\begin{equation}
\delta_i=h.
\label{eq:delta_i_constraint}
\end{equation}
Within the local regime, maximizing $P_S(\boldsymbol{\delta})$ is equivalent
to minimizing
\begin{equation}
\boldsymbol{\delta}^T
\mathbf{J}_{\boldsymbol{\omega}}^{\mathrm{eff}}
\boldsymbol{\delta}
\qquad
\text{subject to }\delta_i=h.
\label{eq:local_quad_form}
\end{equation}
Since
\begin{equation}
\mathbf{C}_{\boldsymbol{\omega}}
=
(\mathbf{J}_{\boldsymbol{\omega}}^{\mathrm{eff}})^{-1},
\label{eq:Comega_recall_sec4}
\end{equation}
the maximizing local direction is
\begin{equation}
\boldsymbol{\delta}_i^\star(h)
=
\frac{
h\,\mathbf{C}_{\boldsymbol{\omega}}\boldsymbol{e}_i
}{
\boldsymbol{e}_i^T
\mathbf{C}_{\boldsymbol{\omega}}
\boldsymbol{e}_i
},
\label{eq:delta_i_star}
\end{equation}
for which
\begin{equation}
(\boldsymbol{\delta}_i^\star)^T
\mathbf{J}_{\boldsymbol{\omega}}^{\mathrm{eff}}
\boldsymbol{\delta}_i^\star
=
\frac{h^2}{
\boldsymbol{e}_i^T
\mathbf{C}_{\boldsymbol{\omega}}
\boldsymbol{e}_i
}.
\label{eq:quad_at_opt}
\end{equation}
Hence
\begin{equation}
P_{S,i}^\star(h)
=
Q\!\left(
\frac{h}{
2\sqrt{
\boldsymbol{e}_i^T
\mathbf{C}_{\boldsymbol{\omega}}
\boldsymbol{e}_i}}
\right).
\label{eq:PSi_star}
\end{equation}

Let
\begin{equation}
\bar{P}_L
=
Q(\bar{\gamma}_L).
\label{eq:PLbar_def}
\end{equation}
Then the crossing rule \eqref{eq:crossing_rule_def} gives
\begin{equation}
\tilde{h}_i
=
2\bar{\gamma}_L
\sqrt{
\boldsymbol{e}_i^T
\mathbf{C}_{\boldsymbol{\omega}}
\boldsymbol{e}_i }.
\label{eq:hi_tilde}
\end{equation}
Substituting \eqref{eq:PSi_star} and \eqref{eq:hi_tilde} into
\eqref{eq:BiL_def} yields
\begin{equation}
B_{i,\mathrm{L}}
=
\int_0^{\tilde h_i}
h\,
\bigl[
Q\!\left(
\frac{h}{
2\sqrt{
\boldsymbol{e}_i^T
\mathbf{C}_{\boldsymbol{\omega}}
\boldsymbol{e}_i}}
\right)
-
Q(\bar{\gamma}_L)
\bigr]dh.
\label{eq:BiL_start}
\end{equation}
With the change of variable
\begin{equation}
u
=
\frac{h}{
2\sqrt{
\boldsymbol{e}_i^T
\mathbf{C}_{\boldsymbol{\omega}}
\boldsymbol{e}_i
}},
\label{eq:u_change_local}
\end{equation}
so that $u\in[0,\bar\gamma_L]$, we obtain
\begin{equation}
B_{i,\mathrm{L}}
=
4
\bigl(
\boldsymbol{e}_i^T
\mathbf{C}_{\boldsymbol{\omega}}
\boldsymbol{e}_i
\bigr)
\int_0^{\bar{\gamma}_L}
u\,
\bigl[
Q(u)-Q(\bar{\gamma}_L)
\bigr]du.
\label{eq:BiL_u_form}
\end{equation}
Using
\begin{align}
&\int_0^a u\,du=\frac{a^2}{2},
\\
&\int_0^a uQ(u)\,du
=
\frac{a^2}{2}Q(a)-\frac{a}{2}\phi(a)+\frac{1}{2}\Phi(a)-\frac{1}{4},
\label{eq:local_integral_identities}
\end{align}
gives
\begin{equation}
B_{i,\mathrm{L}}
=
\alpha_C(\bar{\gamma}_L)\,
\boldsymbol{e}_i^T
\mathbf{C}_{\boldsymbol{\omega}}
\boldsymbol{e}_i,
\label{eq:BiL_final}
\end{equation}
where
\begin{equation}
\alpha_C(a)
=
2a^2Q(a)-2a\phi(a)+2\Phi(a)-1.
\label{eq:alphaC_def_sec4}
\end{equation}
Thus, the local term is available in closed form and provides the
Fisher-consistent component of the basic frequency benchmark.

\subsection{Nonlocal Contribution}
\label{subsec:nonlocal_contribution}

We next construct the nonlocal contribution, whose role is to capture the
prior-scale behavior beyond the local curvature regime. In the nonlocal
region, the relevant mechanism is weak coupling between the two Vandermonde
subspaces. For two steering vectors,
\begin{equation}
\mathbf{a}^H(\omega')\mathbf{a}(\omega)
=
e^{j\frac{(M-1)(\omega-\omega')}{2}}
\frac{
\sin\!\bigl(\frac{M}{2}(\omega-\omega')\bigr)
}{
\sin\!\bigl(\frac{1}{2}(\omega-\omega')\bigr)
},
\label{eq:dirichlet_inner_product}
\end{equation}
so the normalized inner product
\begin{equation}
\frac{1}{M}\mathbf{a}^H(\omega')\mathbf{a}(\omega)
=
e^{j\frac{(M-1)(\omega-\omega')}{2}}
\frac{
\sin\!\bigl(\frac{M}{2}(\omega-\omega')\bigr)
}{
M\sin\!\bigl(\frac{1}{2}(\omega-\omega')\bigr)
}
\label{eq:normalized_dirichlet_inner_product}
\end{equation}
becomes small once $|\omega-\omega'|$ exceeds the main-lobe scale $O(1/M)$.
Accordingly, when every frequency in
$\boldsymbol{\phi}+\boldsymbol{\delta}$ is separated from those in
$\boldsymbol{\phi}$ beyond the local resolution scale,
\begin{equation}
\mathbf{A}^H(\boldsymbol{\phi}+\boldsymbol{\delta})
\mathbf{A}(\boldsymbol{\phi})
\approx
\mathbf{0},
\label{eq:cross_gram_small}
\end{equation}
while the self-Gram matrices remain approximately diagonal:
\begin{equation}
\mathbf{A}^H(\boldsymbol{\phi})\mathbf{A}(\boldsymbol{\phi})
\approx
M\mathbf{I}_K,
\quad
\mathbf{A}^H(\boldsymbol{\phi}+\boldsymbol{\delta})
\mathbf{A}(\boldsymbol{\phi}+\boldsymbol{\delta})
\approx
M\mathbf{I}_K.
\label{eq:self_gram_diag}
\end{equation}
These approximations are used only to construct the representative nonlocal
plateau entering the explicit benchmark.

Under the weak-coupling approximations in \eqref{eq:cross_gram_small} and
\eqref{eq:self_gram_diag}, the nonlocal detector statistics reduce to
\begin{equation}
\mathbf{P}_{\boldsymbol{\phi}+\boldsymbol{\delta}}
\mathbf{A}(\boldsymbol{\phi})
\approx
\mathbf{0},
\qquad
\mathbf{P}_{\boldsymbol{\phi}}
\mathbf{P}_{\boldsymbol{\phi}+\boldsymbol{\delta}}
\approx
\mathbf{0},
\label{eq:projector_weak_coupling}
\end{equation}
and hence
\begin{equation}
\mathbf{P}_{\boldsymbol{\phi}+\boldsymbol{\delta}}\mathbf{M}_0
\approx
\mathbf{0}.
\label{eq:Pnu_M0_small}
\end{equation}
Therefore,
\begin{equation}
\boldsymbol{\Delta}\mathbf{M}_0
=
\bigl(
\mathbf{P}_{\boldsymbol{\phi}}-\mathbf{P}_{\boldsymbol{\phi}+\boldsymbol{\delta}}
\bigr)\mathbf{M}_0
\approx
\mathbf{M}_0,
\label{eq:DeltaM0_nonlocal}
\end{equation}
and
\begin{equation}
\boldsymbol{\Delta}^2
=
\bigl(
\mathbf{P}_{\boldsymbol{\phi}}-\mathbf{P}_{\boldsymbol{\phi}+\boldsymbol{\delta}}
\bigr)^2
\approx
\mathbf{P}_{\boldsymbol{\phi}}
+
\mathbf{P}_{\boldsymbol{\phi}+\boldsymbol{\delta}}.
\label{eq:Delta2_nonlocal}
\end{equation}
Substituting these relations into \eqref{eq:muGamma} and
\eqref{eq:sigmaGamma} gives
\begin{equation}
\mu_\Gamma(\boldsymbol{\phi},\boldsymbol{\delta})
\approx
\|\mathbf{M}_0\|_F^2,
\qquad
\sigma_\Gamma^2(\boldsymbol{\phi},\boldsymbol{\delta})
\approx
2\sigma^2\|\mathbf{M}_0\|_F^2
+
2KT\sigma^4,
\label{eq:nonlocal_mu_sigma}
\end{equation}
since
\begin{equation}
\operatorname{Tr}(\mathbf{P}_{\boldsymbol{\phi}})
=
\operatorname{Tr}(\mathbf{P}_{\boldsymbol{\phi}+\boldsymbol{\delta}})
=
K.
\label{eq:trace_projector_K}
\end{equation}
The corresponding nonlocal score is therefore
\begin{equation}
\gamma_L(\boldsymbol{\phi})
=
\frac{
\|\mathbf{M}_0\|_F^2
}{
\sqrt{
2\sigma^2\|\mathbf{M}_0\|_F^2+2KT\sigma^4
}
},
\label{eq:gammaL_phi}
\end{equation}
with plateau level
\begin{equation}
P_L(\boldsymbol{\phi})
=
Q\!\bigl(\gamma_L(\boldsymbol{\phi})\bigr).
\label{eq:PL_phi}
\end{equation}

To remove the residual dependence on $\boldsymbol{\phi}$, we replace the
self-Gram term by its well-separated approximation in \eqref{eq:self_gram_diag},
which yields
\begin{equation}
\|\mathbf{M}_0\|_F^2
=
\operatorname{Tr}
\!\left(
\mathbf{X}^H
\mathbf{A}^H(\boldsymbol{\phi})
\mathbf{A}(\boldsymbol{\phi})
\mathbf{X}
\right)
\approx
M\|\mathbf{X}\|_F^2.
\label{eq:M0_energy_approx}
\end{equation}
This leads to the representative nonlocal score
\begin{equation}
\bar{\gamma}_L
=
\frac{
M\|\mathbf{X}\|_F^2
}{
\sqrt{
2M\sigma^2\|\mathbf{X}\|_F^2
+
2KT\sigma^4
}
},
\label{eq:gammaL_bar}
\end{equation}
and the representative plateau level
\begin{equation}
\bar{P}_L
=
Q(\bar{\gamma}_L).
\label{eq:PLbar_again}
\end{equation}

Substituting \eqref{eq:PLbar_again} into \eqref{eq:BiNL_def} gives
\begin{equation}
B_{i,\mathrm{NL}}
=
\frac{\bar{P}_L}{2}(I_{i,1}+I_{i,2}),
\label{eq:BiNL_I12}
\end{equation}
where
\begin{align}
&I_{i,1}
=
\int_0^\zeta
h
\int_0^{\zeta-h}
f_i(u)\,du\,dh, \label{eq:Ii_splita}
\\
&I_{i,2}
=
\int_0^\zeta
h
\int_0^{\zeta-h}
f_i(u+h)\,du\,dh.
\label{eq:Ii_splitb}
\end{align}
Exchanging the order of integration yields
\begin{equation}
I_{i,1}
=
\frac{1}{2}
\int_0^\zeta
(\zeta-u)^2f_i(u)\,du
=
\frac{1}{2}\mathbb{E}\!\left[(\zeta-\theta_i)^2\right],
\label{eq:Ii1_eval}
\end{equation}
and, with the change of variable $v=u+h$,
\begin{equation}
I_{i,2}
=
\frac{1}{2}
\int_0^\zeta
v^2f_i(v)\,dv
=
\frac{1}{2}\mathbb{E}\!\left[\theta_i^2\right],
\label{eq:Ii2_eval}
\end{equation}
where $\theta_i$ denotes the $i$th order statistic. Hence
\begin{equation}
B_{i,\mathrm{NL}}
=
\frac{\bar{P}_L}{4}
\left(
\mathbb{E}\!\left[(\zeta-\theta_i)^2\right]
+
\mathbb{E}\!\left[\theta_i^2\right]
\right).
\label{eq:BiNL_moment_form}
\end{equation}
Using
\begin{equation}
\mathbb{E}[\theta_i]
=
\frac{i\zeta}{K+1},
\qquad
\mathbb{E}[\theta_i^2]
=
\frac{i(i+1)}{(K+1)(K+2)}\zeta^2,
\label{eq:orderstat_moments}
\end{equation}
together with
\begin{equation}
\mathbb{E}\!\left[(\zeta-\theta_i)^2\right]
=
\zeta^2-2\zeta\mathbb{E}[\theta_i]+\mathbb{E}[\theta_i^2],
\label{eq:zeta_minus_theta_sq}
\end{equation}
we obtain
\begin{equation}
B_{i,\mathrm{NL}}
=
\frac{\bar{P}_L\zeta^2}{4}
\left[
1-\frac{2i}{K+1}
+
\frac{2i(i+1)}{(K+1)(K+2)}
\right].
\label{eq:BiNL_final}
\end{equation}
Thus, the nonlocal term is also available in closed form and provides the
prior-scale component of the basic frequency benchmark.

\subsection{Coordinate-Wise Basic Frequency Benchmark and Averaging}
\label{subsec:coordwise_basic_benchmark}

Combining \eqref{eq:BiL_final} and \eqref{eq:BiNL_final} gives the
coordinate-wise basic frequency benchmark
\begin{equation}
\boldsymbol{e}_i^T
\mathbf{R}_{\boldsymbol{\omega}}
\boldsymbol{e}_i
\gtrsim
B_{i,\mathrm{L}}+B_{i,\mathrm{NL}},
\label{eq:coordwise_total}
\end{equation}
namely,
\begin{align}
\boldsymbol{e}_i^T
\mathbf{R}_{\boldsymbol{\omega}}
\boldsymbol{e}_i
\gtrsim\;&
\frac{Q(\bar{\gamma}_L)\zeta^2}{4}
\left[
1-\frac{2i}{K+1}
+
\frac{2i(i+1)}{(K+1)(K+2)}
\right]
\nonumber\\
&+
\alpha_C(\bar{\gamma}_L)\,
[\mathbf{C}_{\boldsymbol{\omega}}]_{ii}.
\label{eq:coordwise_total_explicit}
\end{align}

Averaging \eqref{eq:coordwise_total_explicit} over $i=1,\ldots,K$ yields
\begin{equation}
\mathrm{MSE}_{\boldsymbol{\omega}}
=
\frac{1}{K}\sum_{i=1}^K
\boldsymbol{e}_i^T
\mathbf{R}_{\boldsymbol{\omega}}
\boldsymbol{e}_i
\gtrsim
\frac{1}{K}\sum_{i=1}^K
\bigl(B_{i,\mathrm{L}}+B_{i,\mathrm{NL}}\bigr).
\label{eq:MSEomega_avg_start}
\end{equation}
For the local term,
\begin{equation}
\frac{1}{K}\sum_{i=1}^K B_{i,\mathrm{L}}
=
\alpha_C(\bar{\gamma}_L)\,
\frac{1}{K}\operatorname{Tr}(\mathbf{C}_{\boldsymbol{\omega}}).
\label{eq:avg_local_term}
\end{equation}
For the nonlocal term,
\begin{equation}
\frac{1}{K}\sum_{i=1}^K B_{i,\mathrm{NL}}
=
\frac{Q(\bar{\gamma}_L)\zeta^2}{4K}
\sum_{i=1}^K
\left[
1-\frac{2i}{K+1}
+
\frac{2i(i+1)}{(K+1)(K+2)}
\right].
\label{eq:avg_nonlocal_start}
\end{equation}
Using
\begin{align}
&\sum_{i=1}^K 1 = K,
\qquad
\sum_{i=1}^K i = \frac{K(K+1)}{2},
\label{eq:sum_identities_sec4a}
\\
&\sum_{i=1}^K i(i+1)
=
\frac{K(K+1)(K+2)}{3},
\label{eq:sum_identities_sec4b}
\end{align}
we obtain
\begin{equation}
\frac{1}{K}\sum_{i=1}^K B_{i,\mathrm{NL}}
=
Q(\bar{\gamma}_L)\frac{\zeta^2}{6}.
\label{eq:avg_nonlocal_final}
\end{equation}
Hence
\begin{equation}
\mathrm{MSE}_{\boldsymbol{\omega}}
\gtrsim
B_{\boldsymbol{\omega}}^{\mathrm{basic}},
\label{eq:basic_bound_start}
\end{equation}
with
\begin{equation}
B_{\boldsymbol{\omega}}^{\mathrm{basic}}
=
Q(\bar{\gamma}_L)\frac{\zeta^2}{6}
+
\alpha_C(\bar{\gamma}_L)\,
\frac{1}{K}\operatorname{Tr}(\mathbf{C}_{\boldsymbol{\omega}}).
\label{eq:basic_bound_final}
\end{equation}
Equation~\eqref{eq:basic_bound_final} gives the basic explicit benchmark on
the frequency side.

%
\section{Ordered-Prior Correction and Asymptotic Interpretation}
\label{sec:ordered_prior_correction}

Section~\ref{sec:componentwise_zzb} yields the basic frequency benchmark
\begin{equation}
\mathrm{MSE}_{\boldsymbol{\omega}}
\gtrsim
B_{\boldsymbol{\omega}}^{\mathrm{basic}}
=
Q(\bar{\gamma}_L)\frac{\zeta^2}{6}
+
\alpha_C(\bar{\gamma}_L)\,
\frac{1}{K}\operatorname{Tr}(\mathbf{C}_{\boldsymbol{\omega}}).
\label{eq:basic_bound_repeat}
\end{equation}
Its local term is asymptotically consistent with the marginalized
frequency-side CRB, whereas its prior-scale term does not recover the correct
ordered low-SNR limit. This section identifies that mismatch, derives the
ordered a priori bound (APB), and corrects only the prior-scale part of the
basic benchmark.

\subsection{Failure of the Basic A Priori Limit}

We first examine the low-SNR limit of \eqref{eq:basic_bound_repeat}. From
\eqref{eq:gammaL_bar},
\begin{equation}
\bar{\gamma}_L \to 0
\qquad \text{as SNR}\to 0.
\label{eq:gammaL_to_zero}
\end{equation}
Using
\begin{equation}
Q(0)=\frac{1}{2},
\qquad
\Phi(0)=\frac{1}{2},
\qquad
\phi(0)=\frac{1}{\sqrt{2\pi}},
\label{eq:QPhi_zero}
\end{equation}
together with \eqref{eq:alphaC_def_sec4}, we obtain
\begin{equation}
Q(\bar{\gamma}_L)\to \frac{1}{2},
\qquad
\alpha_C(\bar{\gamma}_L)\to 0.
\label{eq:low_snr_limits_basic}
\end{equation}
Hence
\begin{equation}
B_{\boldsymbol{\omega}}^{\mathrm{basic}}
\to
\frac{\zeta^2}{12}.
\label{eq:basic_apb_limit}
\end{equation}

The limit in \eqref{eq:basic_apb_limit} is the scalar a priori bound for an
unsorted uniform parameter over an interval of width $\zeta$, not the correct
prior limit for the ordered frequency vector considered here.\footnote{For
example, if $K=2$ and the true ordered vector is
$[\omega_1,\omega_2]^T$ with $\omega_1<\omega_2$, then the estimate
$[\omega_2,\omega_1]^T$ incurs squared error
$2(\omega_2-\omega_1)^2$ if ordering is ignored, but zero error after sorting.
The relevant MSE geometry is therefore intrinsically tied to the ordered
parameterization.} In the no-information regime, the relevant Bayes risk is
therefore determined by the ordered prior geometry rather than by the variance
of an unsorted scalar parameter.

This mismatch is a consequence of the construction route of
$B_{\boldsymbol{\omega}}^{\mathrm{basic}}$. In
Section~\ref{sec:componentwise_zzb}, the nonlocal contribution is obtained by
forming coordinate-wise terms and then averaging them. This yields a tractable
trace-level expression, but it does not fully preserve the dependence
structure induced by the ordering constraint in the a priori region.
Consequently, the basic benchmark has the correct local limit but an overly
conservative prior-scale limit. The local Fisher-consistent term is not at
issue; only the nonlocal term requires correction.

The correct prior limit is obtained from the Bayes risk of the order
statistics. Let
\begin{equation}
\theta_i \triangleq \omega_{(i)},
\qquad i=1,\ldots,K.
\label{eq:thetai_def}
\end{equation}
For a uniform prior over an interval of width $\zeta$,
\begin{equation}
\mathbb{E}[\theta_i]
=
\frac{i\zeta}{K+1},
\qquad
\operatorname{Var}(\theta_i)
=
\frac{\zeta^2\,i(K+1-i)}{(K+1)^2(K+2)}.
\label{eq:thetai_mean_var}
\end{equation}
Therefore, the ordered APB is
\begin{equation}
\mathrm{APB}_{\mathrm{ord}}
=
\frac{1}{K}
\sum_{i=1}^K
\operatorname{Var}(\theta_i).
\label{eq:ordered_apb_start}
\end{equation}
Substituting \eqref{eq:thetai_mean_var} into \eqref{eq:ordered_apb_start}
gives
\begin{equation}
\mathrm{APB}_{\mathrm{ord}}
=
\frac{\zeta^2}{K(K+1)^2(K+2)}
\sum_{i=1}^K i(K+1-i).
\label{eq:ordered_apb_sub}
\end{equation}
Using
\begin{align}
\sum_{i=1}^K i(K+1-i)
&=
(K+1)\sum_{i=1}^K i-\sum_{i=1}^K i^2
\nonumber\\
&=
(K+1)\frac{K(K+1)}{2}
-
\frac{K(K+1)(2K+1)}{6}
\nonumber\\
&=
\frac{K(K+1)(K+2)}{6},
\label{eq:sum_identity_order}
\end{align}
we obtain
\begin{equation}
\mathrm{APB}_{\mathrm{ord}}
=
\frac{\zeta^2}{6(K+1)}.
\label{eq:ordered_apb_final}
\end{equation}
Equation~\eqref{eq:ordered_apb_final} gives the correct low-SNR limit for the
ordered frequency MSE.

\subsection{Ordered-Prior-Corrected Frequency Benchmark}

The comparison of \eqref{eq:basic_apb_limit} and
\eqref{eq:ordered_apb_final} shows that only the nonlocal term of
\eqref{eq:basic_bound_repeat} must be modified. The local term already has the
desired high-SNR Fisher-consistent behavior and is therefore retained
unchanged.

Recall that the representative nonlocal plateau is
\begin{equation}
\bar{P}_L=Q(\bar{\gamma}_L).
\label{eq:PLbar_repeat}
\end{equation}
In the low-SNR regime, $\bar{\gamma}_L\to 0$ and hence
$2Q(\bar{\gamma}_L)\to 1$. The switching factor therefore already has the
correct limiting behavior; the mismatch lies only in the prior-scale constant.
Accordingly, we retain the same nonlocal switching coefficient and replace the
conservative prior scale $\zeta^2/12$ by the exact ordered APB
$\zeta^2/[6(K+1)]$. We therefore define the corrected nonlocal term as
\begin{equation}
B_{\boldsymbol{\omega},\mathrm{NL}}^{\mathrm{corr}}
=
2Q(\bar{\gamma}_L)\,
\frac{\zeta^2}{6(K+1)}.
\label{eq:nonlocal_corr}
\end{equation}
The local term remains
\begin{equation}
B_{\boldsymbol{\omega},\mathrm{L}}
=
\alpha_C(\bar{\gamma}_L)\,
\frac{1}{K}\operatorname{Tr}(\mathbf{C}_{\boldsymbol{\omega}}).
\label{eq:local_keep}
\end{equation}
Combining the two terms gives the ordered-prior-corrected frequency benchmark
\begin{equation}
B_{\boldsymbol{\omega}}^{\mathrm{corr}}
=
2Q(\bar{\gamma}_L)\,
\frac{\zeta^2}{6(K+1)}
+
\alpha_C(\bar{\gamma}_L)\,
\frac{1}{K}\operatorname{Tr}(\mathbf{C}_{\boldsymbol{\omega}}).
\label{eq:corrected_bound_final}
\end{equation}
Thus, the correction acts only on the prior-scale contribution distorted by
the coordinate-wise averaging route and leaves the local Fisher-consistent
part unchanged.

\subsection{Asymptotic Interpretation}

The corrected benchmark has the desired behavior at both SNR extremes. In the
low-SNR regime, \eqref{eq:gammaL_to_zero} and \eqref{eq:low_snr_limits_basic}
imply
\begin{equation}
2Q(\bar{\gamma}_L)\to 1,
\qquad
\alpha_C(\bar{\gamma}_L)\to 0.
\label{eq:low_snr_corrected}
\end{equation}
Substituting these limits into \eqref{eq:corrected_bound_final} yields
\begin{equation}
B_{\boldsymbol{\omega}}^{\mathrm{corr}}
\to
\frac{\zeta^2}{6(K+1)}
=
\mathrm{APB}_{\mathrm{ord}}.
\label{eq:corrected_low_limit}
\end{equation}
Thus, the corrected benchmark exactly recovers the ordered APB in the
no-information regime.

In the high-SNR regime, \eqref{eq:gammaL_bar} implies
\begin{equation}
\bar{\gamma}_L\to\infty.
\label{eq:gammaL_to_inf}
\end{equation}
Using
\begin{equation}
Q(\bar{\gamma}_L)\to 0,
\qquad
\Phi(\bar{\gamma}_L)\to 1,
\qquad
\bar{\gamma}_L\phi(\bar{\gamma}_L)\to 0,
\label{eq:high_snr_standard_limits}
\end{equation}
it follows from \eqref{eq:alphaC_def_sec4} that
\begin{equation}
\alpha_C(\bar{\gamma}_L)\to 1.
\label{eq:alpha_to_one}
\end{equation}
At the same time,
\begin{equation}
2Q(\bar{\gamma}_L)\,
\frac{\zeta^2}{6(K+1)}
\to 0.
\label{eq:nonlocal_high_limit}
\end{equation}
Therefore,
\begin{equation}
B_{\boldsymbol{\omega}}^{\mathrm{corr}}
\to
\frac{1}{K}\operatorname{Tr}(\mathbf{C}_{\boldsymbol{\omega}}),
\label{eq:corrected_high_limit}
\end{equation}
which is precisely the marginalized frequency-side CRB in trace form.

The corrected benchmark can therefore be written as
\begin{equation}
B_{\boldsymbol{\omega}}^{\mathrm{corr}}
=
\underbrace{
2Q(\bar{\gamma}_L)\,
\frac{\zeta^2}{6(K+1)}
}_{\text{ordered-prior term}}
+
\underbrace{
\alpha_C(\bar{\gamma}_L)\,
\frac{1}{K}\operatorname{Tr}(\mathbf{C}_{\boldsymbol{\omega}})
}_{\text{local Fisher term}}.
\label{eq:corrected_decomposition}
\end{equation}
The first term dominates in the ordered a priori region, whereas the second
term dominates in the asymptotic region. The correction therefore has a
single role: it replaces the conservative prior-scale contribution in
\eqref{eq:basic_bound_repeat} by the correct ordered-prior term while
preserving the local Fisher-consistent structure of the basic benchmark.
\section{Local Transfer Analysis of the Plug-In Amplitude MSE}
\label{sec:induced_x_benchmark}

This section turns to the amplitude side. Unlike the frequency-side analysis,
which is global and ZZB-type, the quantity of practical interest here is the
MSE of the sequential plug-in estimator induced by frequency uncertainty.
Accordingly, we develop a local transfer characterization for the plug-in
amplitude reconstruction error and then derive a computable benchmark for the
medium-to-high-SNR regime.

More specifically, the exact plug-in MSE
\begin{equation}
\mathrm{MSE}_{\mathbf{X}}
=
\frac{1}{KT}
\mathbb{E}
\!\left[
\|\hat{\mathbf{X}}(\hat{\boldsymbol{\omega}})-\mathbf{X}\|_F^2
\right]
\label{eq:MSEX_exact_def}
\end{equation}
does not in general admit a simple closed-form expression, because it couples
the nonlinear frequency estimation error, the additive noise, and the
reconstruction map. In the regime where
$\hat{\boldsymbol{\omega}}$ remains in a neighborhood of the true frequency
vector, however, the induced amplitude error can be described through a local
second-order transfer law. The analysis below is therefore local; it is not a
second global low-SNR characterization parallel to the frequency-side ZZB-type
benchmark.

\subsection{Oracle Baseline and Local Transfer Form}
\label{subsec:X_oracle_transfer}

We begin with the oracle reconstruction error and then isolate the additional
error induced by frequency mismatch. For any candidate frequency vector
$\boldsymbol{\nu}$, define the plug-in least-squares estimator
\begin{equation}
\hat{\mathbf{X}}(\boldsymbol{\nu})
=
\bigl(
\mathbf{A}^H(\boldsymbol{\nu})\mathbf{A}(\boldsymbol{\nu})
\bigr)^{-1}
\mathbf{A}^H(\boldsymbol{\nu})\mathbf{Y}.
\label{eq:Xhat_nu}
\end{equation}
When $\boldsymbol{\nu}=\boldsymbol{\omega}$, this reduces to the oracle
least-squares estimator. Substituting
$\mathbf{Y}=\mathbf{A}(\boldsymbol{\omega})\mathbf{X}+\mathbf{N}$ gives
\begin{equation}
\hat{\mathbf{X}}(\boldsymbol{\omega})-\mathbf{X}
=
\bigl(
\mathbf{A}^H(\boldsymbol{\omega})\mathbf{A}(\boldsymbol{\omega})
\bigr)^{-1}
\mathbf{A}^H(\boldsymbol{\omega})\mathbf{N}.
\label{eq:Xhat_oracle_error}
\end{equation}
Hence the oracle amplitude baseline is
\begin{equation}
B_{\mathbf{X}}^{\mathrm{or}}
\triangleq
\frac{1}{KT}
\mathbb{E}
\!\left[
\|\hat{\mathbf{X}}(\boldsymbol{\omega})-\mathbf{X}\|_F^2
\right]
=
\frac{\sigma^2}{K}
\operatorname{Tr}
\!\left[
\bigl(
\mathbf{A}^H(\boldsymbol{\omega})\mathbf{A}(\boldsymbol{\omega})
\bigr)^{-1}
\right].
\label{eq:BX_oracle}
\end{equation}
This is the irreducible noise-limited amplitude error under perfect frequency
knowledge.

To isolate the additional error induced by frequency mismatch, let the true
frequency vector be $\boldsymbol{\phi}$ and reconstruct the amplitudes using a
candidate vector $\boldsymbol{\nu}$. Define the transfer matrix
\begin{equation}
\mathbf{T}(\boldsymbol{\phi},\boldsymbol{\nu})
\triangleq
\bigl(
\mathbf{A}^H(\boldsymbol{\nu})\mathbf{A}(\boldsymbol{\nu})
\bigr)^{-1}
\mathbf{A}^H(\boldsymbol{\nu})\mathbf{A}(\boldsymbol{\phi}),
\label{eq:T_def}
\end{equation}
and the corresponding deterministic mismatch kernel
\begin{equation}
d_{\mathbf{X}}^2(\boldsymbol{\phi},\boldsymbol{\nu})
\triangleq
\frac{1}{KT}
\left\|
\bigl(
\mathbf{I}_K-\mathbf{T}(\boldsymbol{\phi},\boldsymbol{\nu})
\bigr)\mathbf{X}
\right\|_F^2.
\label{eq:dX_def}
\end{equation}

Now let $\boldsymbol{\nu}=\boldsymbol{\phi}+\boldsymbol{\delta}$. Since
\begin{equation}
\mathbf{T}(\boldsymbol{\phi},\boldsymbol{\phi})=\mathbf{I}_K,
\label{eq:T_identity}
\end{equation}
one has
\begin{equation}
d_{\mathbf{X}}^2(\boldsymbol{\phi},\boldsymbol{\phi})=0,
\label{eq:dX_zero}
\end{equation}
so the first nonzero term of
$d_{\mathbf{X}}^2(\boldsymbol{\phi},\boldsymbol{\phi}+\boldsymbol{\delta})$
is second-order in $\boldsymbol{\delta}$. This gives the local expansion
\begin{equation}
d_{\mathbf{X}}^2(\boldsymbol{\phi},\boldsymbol{\phi}+\boldsymbol{\delta})
=
\boldsymbol{\delta}^T
\mathbf{H}_{\mathbf{X}}(\boldsymbol{\phi})
\boldsymbol{\delta}
+
o(\|\boldsymbol{\delta}\|^2),
\label{eq:dX_local_hessian}
\end{equation}
where
\begin{align}
[\mathbf{H}_{\mathbf{X}}(\boldsymbol{\phi})]_{ij}
=&
\frac{1}{KT}
\Re\Bigg\{
\operatorname{Tr}
\Bigg(
\mathbf{X}^H
\mathbf{A}_i^H(\boldsymbol{\phi})
\mathbf{A}(\boldsymbol{\phi}) \nonumber \\
&\times
\bigl(
\mathbf{A}^H(\boldsymbol{\phi})\mathbf{A}(\boldsymbol{\phi})
\bigr)^{-2}
\mathbf{A}^H(\boldsymbol{\phi})
\mathbf{A}_j(\boldsymbol{\phi})
\mathbf{X}
\Bigg)
\Bigg\},
\label{eq:HX_trace_form}
\end{align}
with
\begin{equation}
\mathbf{A}_i(\boldsymbol{\phi})
=
\frac{\partial \mathbf{A}(\boldsymbol{\phi})}{\partial \phi_i}.
\label{eq:Ai_def}
\end{equation}

Let
\begin{equation}
\boldsymbol{\delta}
=
\hat{\boldsymbol{\omega}}-\boldsymbol{\omega},
\qquad
\mathbf{R}_{\boldsymbol{\omega}}
=
\mathbb{E}
\!\left[
\boldsymbol{\delta}\boldsymbol{\delta}^T
\right].
\label{eq:Romega_def_sec6}
\end{equation}
Then, in the local regime,
\begin{equation}
\mathbb{E}
\!\left[
d_{\mathbf{X}}^2(\boldsymbol{\omega},\hat{\boldsymbol{\omega}})
\right]
=
\operatorname{Tr}
\!\left(
\mathbf{H}_{\mathbf{X}}(\boldsymbol{\omega})
\mathbf{R}_{\boldsymbol{\omega}}
\right)
+
o(\|\mathbf{R}_{\boldsymbol{\omega}}\|),
\label{eq:local_transfer_expectation}
\end{equation}
and hence
\begin{equation}
\mathrm{MSE}_{\mathbf{X}}
=
B_{\mathbf{X}}^{\mathrm{or}}
+
\operatorname{Tr}
\!\left(
\mathbf{H}_{\mathbf{X}}(\boldsymbol{\omega})
\mathbf{R}_{\boldsymbol{\omega}}
\right)
+
o(\|\mathbf{R}_{\boldsymbol{\omega}}\|).
\label{eq:MSEX_local_transfer}
\end{equation}
Equation~\eqref{eq:MSEX_local_transfer} is the basic local transfer relation:
the plug-in amplitude MSE consists of the oracle baseline plus the
second-order propagation of the frequency error covariance through the
reconstruction map.

\subsection{Local Structure of $\mathrm{CRB}_{\mathbf{X}}$}
\label{subsec:CRBX_structure}

We next identify the corresponding local structure of
$\mathrm{CRB}_{\mathbf{X}}$. Recall from Section~\ref{sec:system_model} that
the joint CRB block associated with the real-valued amplitude parameter
$\mathbf{x}$ is
\begin{equation}
\mathbf{C}_{\mathbf{x}}
=
\bigl[
\mathbf{J}_{\boldsymbol{\eta}}^{-1}
\bigr]_{\mathbf{x}\mathbf{x}},
\label{eq:Cx_def_sec6}
\end{equation}
and
\begin{equation}
\mathrm{CRB}_{\mathbf{X}}
=
\frac{1}{KT}\operatorname{Tr}(\mathbf{C}_{\mathbf{x}}).
\label{eq:CRBX_def_sec6}
\end{equation}
By block inversion,
\begin{equation}
\mathbf{C}_{\mathbf{x}}
=
\mathbf{J}_{\mathbf{x}\mathbf{x}}^{-1}
+
\mathbf{J}_{\mathbf{x}\mathbf{x}}^{-1}
\mathbf{J}_{\mathbf{x}\boldsymbol{\omega}}
\mathbf{C}_{\boldsymbol{\omega}}
\mathbf{J}_{\boldsymbol{\omega}\mathbf{x}}
\mathbf{J}_{\mathbf{x}\mathbf{x}}^{-1},
\label{eq:Cx_block_inverse}
\end{equation}
where
\begin{equation}
\mathbf{C}_{\boldsymbol{\omega}}
=
\Bigl(
\mathbf{J}_{\boldsymbol{\omega}\boldsymbol{\omega}}
-
\mathbf{J}_{\boldsymbol{\omega}\mathbf{x}}
\mathbf{J}_{\mathbf{x}\mathbf{x}}^{-1}
\mathbf{J}_{\mathbf{x}\boldsymbol{\omega}}
\Bigr)^{-1}
\label{eq:Comega_repeat_sec6}
\end{equation}
is the marginalized frequency-side CRB.

After normalization by $KT$, the first term in
\eqref{eq:Cx_block_inverse} yields the oracle baseline in
\eqref{eq:BX_oracle}, whereas the second term is the local penalty induced by
frequency uncertainty. Thus, the amplitude-side CRB has the same
oracle-plus-transfer structure as the local plug-in expansion. In the same
local coordinates as \eqref{eq:dX_local_hessian}, one obtains
\begin{equation}
\mathrm{CRB}_{\mathbf{X}}
=
B_{\mathbf{X}}^{\mathrm{or}}
+
\operatorname{Tr}
\!\left(
\mathbf{H}_{\mathbf{X}}(\boldsymbol{\omega})
\mathbf{C}_{\boldsymbol{\omega}}
\right).
\label{eq:CRBX_transfer_form}
\end{equation}

Equations~\eqref{eq:MSEX_local_transfer} and
\eqref{eq:CRBX_transfer_form} show that the local plug-in MSE and the local
amplitude-side CRB differ only in the frequency-side quantity that drives the
transfer: the former depends on the actual frequency error covariance
$\mathbf{R}_{\boldsymbol{\omega}}$, whereas the latter depends on the local
Fisher limit $\mathbf{C}_{\boldsymbol{\omega}}$.

\subsection{Structured Approximation and Computable Benchmark}
\label{subsec:HX_structure}

The relations \eqref{eq:MSEX_local_transfer} and
\eqref{eq:CRBX_transfer_form} are informative but not yet directly
computable, because the full matrix $\mathbf{R}_{\boldsymbol{\omega}}$ is not
available in closed form. The purpose of the following approximation is not to
diagonalize the transfer Hessian exactly, but to reduce the amplitude-side
benchmark to quantities already made explicit on the frequency side.

Write
\begin{equation}
\mathbf{G}(\boldsymbol{\omega})
\triangleq
\mathbf{A}^H(\boldsymbol{\omega})\mathbf{A}(\boldsymbol{\omega}),
\label{eq:Gomega_def}
\end{equation}
and let $\mathbf{x}_i^T$ denote the $i$th row of $\mathbf{X}$. Since
$\mathbf{A}_i(\boldsymbol{\omega})$ has only one nonzero column, namely the
derivative of the $i$th steering vector, we may write
\begin{equation}
\mathbf{A}_i(\boldsymbol{\omega})
=
\mathbf{a}_i'(\omega_i)\mathbf{e}_i^T,
\label{eq:Ai_rankone}
\end{equation}
where
\begin{equation}
\mathbf{a}_i'(\omega_i)
\triangleq
\frac{\partial \mathbf{a}(\omega_i)}{\partial \omega_i}.
\label{eq:ai_prime_def}
\end{equation}
Define
\begin{equation}
\mathbf{p}_i(\boldsymbol{\omega})
\triangleq
\mathbf{G}^{-1}(\boldsymbol{\omega})
\mathbf{A}^H(\boldsymbol{\omega})
\mathbf{a}_i'(\omega_i).
\label{eq:pi_def}
\end{equation}
Then \eqref{eq:HX_trace_form} can be rewritten as
\begin{equation}
[\mathbf{H}_{\mathbf{X}}(\boldsymbol{\omega})]_{ij}
=
\frac{1}{KT}
\Re\!\left\{
(\mathbf{x}_i^H\mathbf{x}_j)
\bigl(
\mathbf{p}_i^H(\boldsymbol{\omega})
\mathbf{p}_j(\boldsymbol{\omega})
\bigr)
\right\}.
\label{eq:HX_compact}
\end{equation}
This form makes the transfer structure explicit: the Hessian is determined by
the interaction between the amplitude-row correlations and the
manifold-derivative correlations.

We next use the same well-separated approximation adopted in
Section~\ref{sec:componentwise_zzb}. When the frequencies are sufficiently
separated,
\begin{equation}
\mathbf{A}^H(\boldsymbol{\omega})\mathbf{A}(\boldsymbol{\omega})
\approx
M\mathbf{I}_K,
\label{eq:gram_approx_sec6}
\end{equation}
and, because the derivative of the $i$th steering vector is nearly orthogonal
to the remaining steering vectors,
\begin{equation}
\mathbf{A}^H(\boldsymbol{\omega})\mathbf{a}_i'(\omega_i)
\approx
j\frac{M(M-1)}{2}\mathbf{e}_i.
\label{eq:Ahaiprime_approx}
\end{equation}
Combining \eqref{eq:gram_approx_sec6} and
\eqref{eq:Ahaiprime_approx} yields
\begin{equation}
\mathbf{p}_i(\boldsymbol{\omega})
\approx
j\frac{M-1}{2}\mathbf{e}_i.
\label{eq:pi_approx}
\end{equation}

Substituting \eqref{eq:pi_approx} into \eqref{eq:HX_compact} gives
\begin{equation}
\mathbf{H}_{\mathbf{X}}(\boldsymbol{\omega})
\approx
\operatorname{Diag}
\!\left(
\frac{(M-1)^2}{4KT}\|\mathbf{x}_1\|_2^2,
\ldots,
\frac{(M-1)^2}{4KT}\|\mathbf{x}_K\|_2^2
\right).
\label{eq:HX_diag_approx}
\end{equation}
This approximation is not an exact diagonalization; it is a structured
simplification that makes the amplitude-side benchmark computable and remains
consistent with the same well-separated Vandermonde approximation used on the
frequency side.

Under \eqref{eq:HX_diag_approx},
\eqref{eq:MSEX_local_transfer} and \eqref{eq:CRBX_transfer_form} reduce to
\begin{equation}
\mathrm{MSE}_{\mathbf{X}}
\approx
B_{\mathbf{X}}^{\mathrm{or}}
+
\frac{(M-1)^2}{4KT}
\sum_{i=1}^K
\|\mathbf{x}_i\|_2^2
[\mathbf{R}_{\boldsymbol{\omega}}]_{ii},
\label{eq:MSEX_diag_form}
\end{equation}
and
\begin{equation}
\mathrm{CRB}_{\mathbf{X}}
\approx
B_{\mathbf{X}}^{\mathrm{or}}
+
\frac{(M-1)^2}{4KT}
\sum_{i=1}^K
\|\mathbf{x}_i\|_2^2
[\mathbf{C}_{\boldsymbol{\omega}}]_{ii}.
\label{eq:CRBX_diag_form}
\end{equation}
Thus, after this approximation, the plug-in amplitude MSE depends only on the
diagonal frequency error levels.

We accordingly adopt the diagonal proxy
\begin{equation}
\widetilde{\mathbf{R}}_{\boldsymbol{\omega}}
=
\operatorname{Diag}
\!\Bigl(
B_{\omega,1}^{\mathrm{basic}},
\ldots,
B_{\omega,K}^{\mathrm{basic}}
\Bigr),
\label{eq:Romega_proxy_sec6}
\end{equation}
where $B_{\omega,i}^{\mathrm{basic}}$ are the coordinate-wise frequency
benchmarks derived in Section~\ref{sec:componentwise_zzb}. This choice is
essential. The correction introduced in
Section~\ref{sec:ordered_prior_correction} is a trace-level correction designed
to restore the proper ordered low-SNR asymptote of the frequency MSE. By
contrast, the local transfer law in \eqref{eq:MSEX_diag_form} depends on
coordinate-wise diagonal error levels. For this reason, the relevant explicit
proxy here is the componentwise basic benchmark
$B_{\omega,i}^{\mathrm{basic}}$, whose role in the present section is local and
coordinate-wise rather than prior-scale and trace-level.

The resulting plug-in amplitude benchmark is
\begin{equation}
B_{\mathbf{X}}^{\mathrm{plug}}
\triangleq
B_{\mathbf{X}}^{\mathrm{or}}
+
\frac{(M-1)^2}{4KT}
\sum_{i=1}^K
\|\mathbf{x}_i\|_2^2
B_{\omega,i}^{\mathrm{basic}}.
\label{eq:BX_plug_final}
\end{equation}
This is a benchmark for the actual sequential plug-in estimator, not another
Fisher-information-based bound. Its practical role is therefore different from
that of the amplitude-side CRB: the latter is a purely local Fisher
reference, whereas $B_{\mathbf{X}}^{\mathrm{plug}}$ is intended to track the
actual sequential plug-in estimator through the frequency-error proxy
inherited from the frequency-side analysis.

Finally, the high-SNR behavior follows directly from the frequency-side
analysis. Since
\begin{equation}
B_{\omega,i}^{\mathrm{basic}}
\to
[\mathbf{C}_{\boldsymbol{\omega}}]_{ii},
\qquad
\mathrm{SNR}\to\infty,
\label{eq:Bi_to_Cii}
\end{equation}
substituting \eqref{eq:Bi_to_Cii} into \eqref{eq:BX_plug_final} gives
\begin{equation}
B_{\mathbf{X}}^{\mathrm{plug}}
\to
\mathrm{CRB}_{\mathbf{X}},
\qquad
\mathrm{SNR}\to\infty,
\label{eq:BX_plug_to_CRB}
\end{equation}
under the same structural approximation. Thus,
$B_{\mathbf{X}}^{\mathrm{plug}}$ inherits the correct local asymptotic limit
while remaining more representative of the actual sequential plug-in estimator
than the pure local amplitude-side CRB in the medium-to-high-SNR regime.
\section{Numerical Results}
\label{sec:simulation}

This section evaluates the proposed benchmarks. We first assess the Gaussian
approximation underlying the GLRT-based kernel used in the explicit ZZB-type
construction, and then examine the frequency-side and amplitude-side behavior
across different SNR regimes, snapshot numbers, and model orders. In all
experiments, the amplitude matrix is deterministic, generated once for each
simulation configuration, and then kept fixed while the noise variance is
adjusted according to the target SNR. On the frequency side, the estimator
under test is MUSIC \cite{wang2017coarrays}; on the amplitude side, the
amplitudes are reconstructed by the maximum likelihood estimator using the estimated frequencies.

\subsection{Validation of the Gaussianized GLRT Kernel}
\label{subsec:sim_glrt_gaussian}

We first assess the Gaussian approximation of the GLRT statistic used in
\eqref{eq:Gamma_gaussian}--\eqref{eq:glrt_kernel_gaussian}. In this
experiment, we use a $K$-tone line spectral model with $K=5$, $M=50$, and
$T=20$. The true ordered frequency vector is
\begin{equation}
\boldsymbol{\phi}
=
[-0.35,\,-0.15,\,0,\,0.15,\,0.35]^T\pi,
\end{equation}
and the deterministic amplitude matrix $\mathbf{X}$ is generated once and then
kept fixed. Two representative perturbations are considered,
\begin{equation}
\boldsymbol{\delta}_{\mathrm{loc}}
=
[0,\;0,\;0.01\pi,\;0,\;0]^T,
\quad
\boldsymbol{\delta}_{\mathrm{mod}}
=
[0,\;0,\;0.10\pi,\;0,\;0]^T.
\end{equation}

\begin{figure}[t]
    \centering
    \includegraphics[width=0.96\columnwidth]{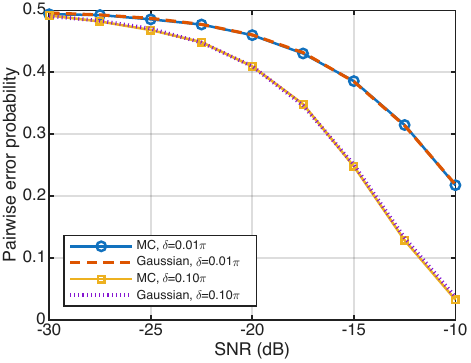}
    \caption{Empirical pairwise error probability of the GLRT statistic and its Gaussian approximation for two representative perturbations.}
    \label{fig:glrt_kernel_validation}
\end{figure}

Fig.~\ref{fig:glrt_kernel_validation} compares the empirical pairwise error
probability
\[
\Pr\!\bigl(\Gamma(\mathbf{Y};\boldsymbol{\phi},\boldsymbol{\delta})<0
\mid \mathcal{H}_0\bigr)
\]
with its Gaussian approximation $Q(\mu_\Gamma/\sigma_\Gamma)$ over the SNR
range of interest. The Monte Carlo and Gaussianized curves are in close
agreement for both perturbations, indicating that the Gaussianized GLRT kernel
captures the pairwise testing behavior relevant to the ZZB-type construction.

\begin{figure}[t]
    \centering
    \includegraphics[width=0.96\columnwidth]{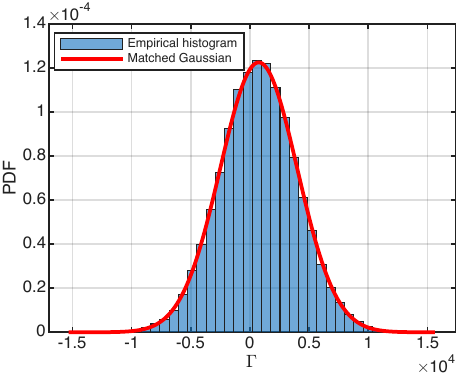}
    \caption{Empirical distribution of the GLRT statistic under $\mathcal{H}_0$ and its matched Gaussian approximation for a representative operating point.}
    \label{fig:glrt_stat_distribution}
\end{figure}

Fig.~\ref{fig:glrt_stat_distribution} shows the empirical distribution of the
GLRT statistic under $\mathcal{H}_0$ for a representative operating point,
together with the Gaussian distribution matched to $\mu_\Gamma$ and
$\sigma_\Gamma^2$. The close fit supports the approximation adopted in
\eqref{eq:Gamma_gaussian}--\eqref{eq:glrt_kernel_gaussian}.

\subsection{Validation of the Switching Coefficients and the Frequency-Side Correction}
\label{subsec:sim_w_validation}

We next validate the switching mechanism underlying the corrected
frequency-side benchmark. In this experiment, we consider a $K$-tone line
spectral model with $K=5$, observation dimension $M=50$, and $T=100$
snapshots. The true ordered frequency vector is
\begin{equation}
\boldsymbol{\omega}
=
[-0.35,\,-0.15,\,0,\,0.15,\,0.35]^T\pi,
\end{equation}
with prior support $[-\pi,\pi]$. The deterministic amplitude matrix
$\mathbf{X}$ is generated once, normalized in Frobenius norm, and then kept
fixed throughout the SNR sweep. For each SNR point, the empirical frequency
MSE of MUSIC is evaluated over $1000$ Monte Carlo runs.

\begin{figure}[t]
    \centering
    \includegraphics[width=0.96\columnwidth]{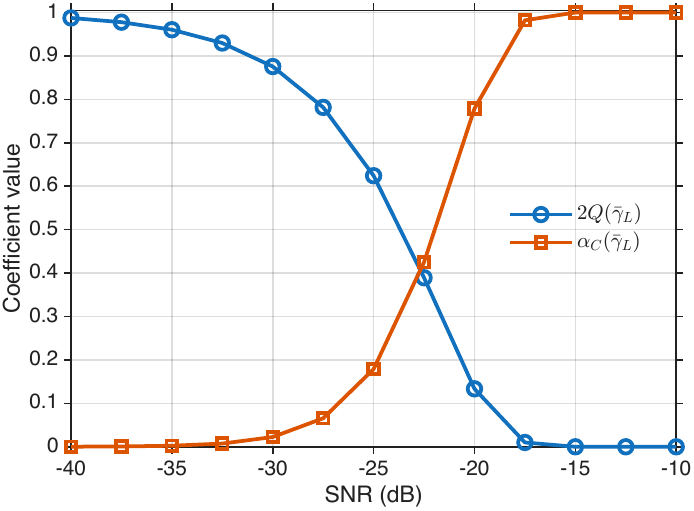}
    \caption{Switching coefficients underlying the ordered-prior-corrected frequency benchmark.}
    \label{fig:q_coefficients}
\end{figure}

Fig.~\ref{fig:q_coefficients} shows the two coefficients
$2Q(\bar{\gamma}_L)$ and $\alpha_C(\bar{\gamma}_L)$ in the corrected
frequency benchmark \eqref{eq:corrected_bound_final}. As the SNR increases,
$2Q(\bar{\gamma}_L)$ decreases smoothly from near one to near zero, whereas
$\alpha_C(\bar{\gamma}_L)$ increases from near zero to near one. Thus, the
corrected benchmark implements a smooth transition from the ordered-prior
regime to the local Fisher regime.

\begin{figure}[t]
    \centering
    \includegraphics[width=0.96\columnwidth]{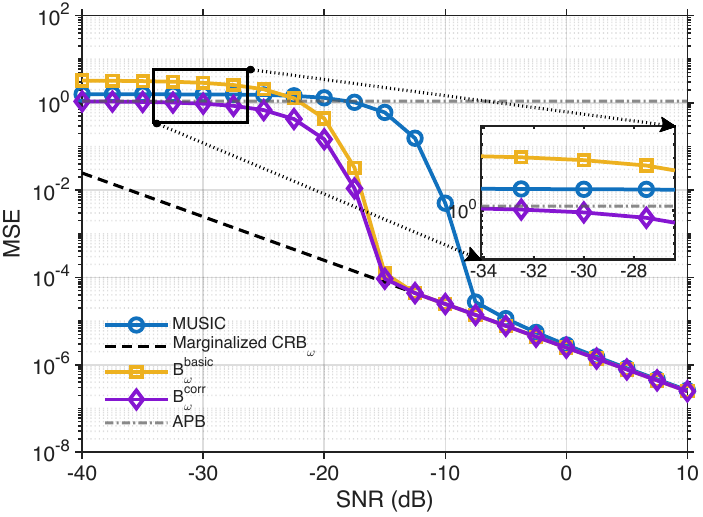}
    \caption{Frequency-side validation of the basic and ordered-prior-corrected benchmarks against MUSIC, the marginalized frequency-side CRB, and the ordered APB.}
    \label{fig:w_bound_comparison}
\end{figure}

Fig.~\ref{fig:w_bound_comparison} compares the empirical MUSIC MSE with the
marginalized frequency-side CRB \eqref{eq:Comega_eff}, the basic benchmark
$B_{\boldsymbol{\omega}}^{\mathrm{basic}}$ in \eqref{eq:basic_bound_final},
the corrected benchmark $B_{\boldsymbol{\omega}}^{\mathrm{corr}}$ in
\eqref{eq:corrected_bound_final}, and the ordered APB
\eqref{eq:ordered_apb_final}. The corrected benchmark approaches the ordered
APB at low SNR, approaches the marginalized CRB at high SNR, and tracks the
transition of the empirical MUSIC curve in the threshold region. By contrast,
the basic benchmark remains overly conservative in the ordered a priori region
and may even exceed the empirical MUSIC MSE, as shown in the inset. This
verifies that the ordered-prior correction is necessary to recover the proper
low-SNR limit without altering the high-SNR Fisher asymptote.

\subsection{Effect of the Number of Snapshots}
\label{subsec:effect_of_T}

We next examine the effect of the number of snapshots. In this experiment, we
fix $M=20$ and $K=3$, and consider
\begin{equation}
T \in \{10,20,40\}.
\end{equation}
The true ordered frequency vector is
\begin{equation}
\boldsymbol{\omega}
=
[-0.3,\;0,\;0.3]^T\pi,
\end{equation}
with prior support $[-0.5\pi,\;0.5\pi]$. For each value of $T$, the empirical
results are averaged over $1000$ Monte Carlo runs.

\begin{figure}[t]
    \centering
    \includegraphics[width=0.98\columnwidth]{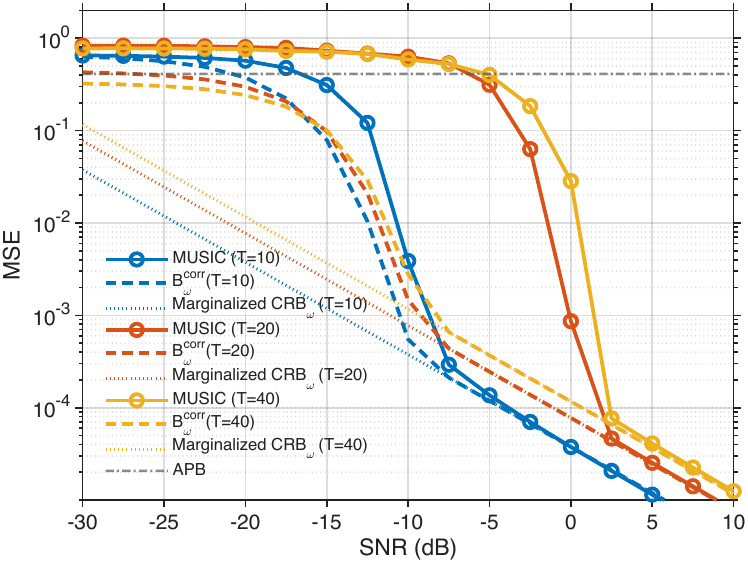}
    \caption{Frequency-side validation of the ordered-prior-corrected frequency benchmark for different numbers of snapshots.}
    \label{fig:w_varying_T}
\end{figure}

Fig.~\ref{fig:w_varying_T} shows the frequency-side results. As $T$
increases, the MUSIC threshold shifts to lower SNR, indicating improved
frequency estimation. The corrected benchmark follows the same trend and
remains anchored by the ordered APB at low SNR and the marginalized
frequency-side CRB at high SNR. Since the ordered APB depends only on the
ordered prior support, it does not vary with $T$.

\begin{figure}[t]
    \centering
    \includegraphics[width=0.98\columnwidth]{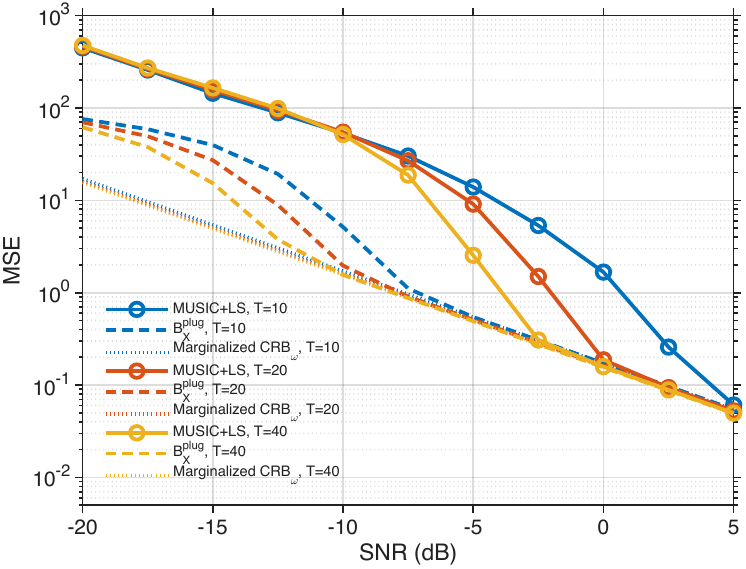}
    \caption{Amplitude-side validation of the plug-in amplitude benchmark and the amplitude-side CRB for different numbers of snapshots.}
    \label{fig:x_varying_T}
\end{figure}

Fig.~\ref{fig:x_varying_T} shows the corresponding amplitude-side results.
As $T$ increases, the empirical MUSIC+LS MSE, the plug-in benchmark
$B_{\mathbf{X}}^{\mathrm{plug}}$, and the amplitude-side CRB all move
downward, showing that the gain on the frequency side is transferred to the
amplitude reconstruction stage. The plug-in benchmark is most accurate in the
moderate-to-high-SNR region, where the local transfer approximation is most
reliable, while still capturing the overall trend at lower SNR.

\subsection{Effect of the Model Order}
\label{subsec:effect_of_K}

Finally, we examine the effect of the model order. We consider
\begin{equation}
K \in \{2,3,4\}.
\end{equation}
For each value of $K$, we generate $10$ independent ordered frequency vectors,
where each component is drawn independently from the uniform distribution on
$[-\pi,\pi]$ and then sorted increasingly. For each such frequency
configuration, one deterministic amplitude matrix $\mathbf{X}$ is generated
and kept fixed throughout the SNR sweep. Under each fixed
$(\boldsymbol{\omega},\mathbf{X})$ pair, $100$ Monte Carlo trials are
performed at each SNR point. The final curves are obtained by averaging first
over the Monte Carlo runs and then over the $10$ random frequency
configurations.

\begin{figure}[t]
    \centering
    \includegraphics[width=0.98\columnwidth]{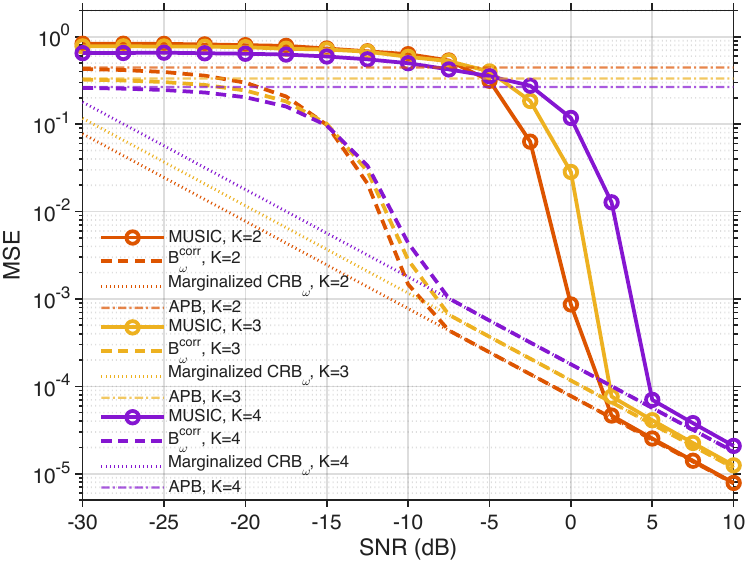}
    \caption{Frequency-side validation of the ordered-prior-corrected frequency benchmark for different model orders.}
    \label{fig:w_varying_K}
\end{figure}

Fig.~\ref{fig:w_varying_K} shows the frequency-side results. As $K$ increases
from $2$ to $4$, the empirical MUSIC MSE shifts upward and to the right,
indicating a clear degradation in threshold behavior. The corrected benchmark
shows the same trend and remains consistent with the ordered APB at low SNR
and the marginalized frequency-side CRB at high SNR.

\begin{figure}[t]
    \centering
    \includegraphics[width=0.98\columnwidth]{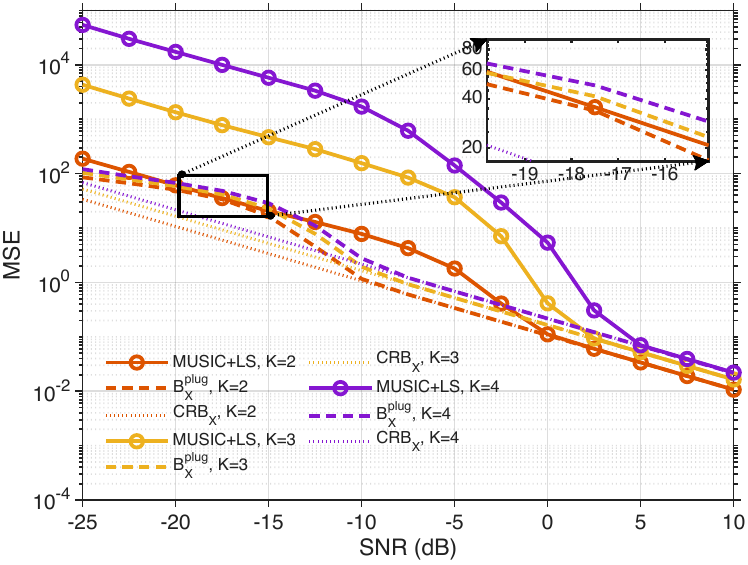}
    \caption{Amplitude-side validation of the plug-in amplitude benchmark and the amplitude-side CRB for different model orders.}
    \label{fig:x_varying_K}
\end{figure}

Fig.~\ref{fig:x_varying_K} shows the amplitude-side results. As $K$
increases, the empirical MUSIC+LS MSE rises substantially, especially in the
low-to-moderate-SNR regime, showing that the increased difficulty of the
frequency estimation stage strongly degrades amplitude reconstruction. The
plug-in benchmark follows the same trend and remains most informative in the
moderate-to-high-SNR region, where the local approximation underlying the
transfer benchmark is most accurate.

\section{Conclusion}
\label{sec:conclusion}

This paper developed explicit performance benchmarks for line spectral
estimation from two complementary perspectives: ordered frequency estimation
and plug-in amplitude reconstruction. On the frequency side, we constructed a
computable ZZB-type benchmark under an ordered prior by using a GLRT-based
surrogate for the unavailable pairwise kernel associated with unknown
deterministic amplitudes. We further introduced an ordered-prior correction so
that the resulting benchmark recovers the ordered APB at low SNR and the
marginalized frequency-side CRB at high SNR. On the amplitude side, we
derived a local transfer characterization for the sequential plug-in
estimator and obtained a computable benchmark for the induced amplitude MSE. The resulting framework separates two coupled aspects of line spectral
estimation performance: threshold behavior in ordered frequency estimation and
error propagation in sequential amplitude reconstruction. Numerical results
showed that the corrected frequency benchmark captures the threshold
transition, whereas the plug-in amplitude benchmark remains informative in the
medium-to-high-SNR regime across different snapshot numbers and model orders. Future work will consider extensions to more general signal models and to
amplitude-side characterizations beyond the local plug-in regime.
\bibliographystyle{IEEEtran}
\bibliography{reference}

%

\end{document}